\documentclass[journal,comsoc]{IEEEtran}
\usepackage[T1]{fontenc}

\usepackage{cite}

\ifCLASSINFOpdf
   \usepackage[pdftex]{graphicx}
\else
\fi
\usepackage{amsmath}
\usepackage{lipsum}
 
\usepackage{xcolor}
\usepackage{amsthm}

\usepackage[cmintegrals]{newtxmath}
\usepackage{algorithm}
\usepackage{booktabs}
\usepackage{subfigure}
\usepackage{tabularx}
\usepackage{algpseudocode}
\usepackage{multirow}
\usepackage{makecell}

\usepackage{array}
\newcolumntype{L}[1]{>{\raggedright\let\newline\\\arraybackslash\hspace{0pt}}m{#1}}
\newcolumntype{C}[1]{>{\centering\let\newline\\\arraybackslash\hspace{0pt}}m{#1}}
\newcolumntype{R}[1]{>{\raggedleft\let\newline\\\arraybackslash\hspace{0pt}}m{#1}}

\usepackage{booktabs}
\usepackage{bbm}
\usepackage{lscape} 

\begin{document}
%
\title{A State-of-the-art Survey on \\Full-duplex Network Design}
%
%
%

\author{Yonghwi~Kim,~\IEEEmembership{Student Member, IEEE,} Hyung-Joo Moon,~\IEEEmembership{Student Member, IEEE,}\\ Hanju Yoo,~\IEEEmembership{Student Member,~IEEE,} Byoungnam (Klaus) Kim~\IEEEmembership{Member,~IEEE,} \\
        Kai-Kit Wong,~\IEEEmembership{Fellow, IEEE}, and Chan-Byoung~Chae,~\IEEEmembership{Fellow, IEEE}
\thanks{Y. Kim, H.-J. Moon, H. Yoo, and C.-B. Chae are with the School of Integrated Technology, Yonsei University, Seoul 03722, South Korea (e-mail: eric\_kim@yonsei.ac.kr; moonhj@yonsei.ac.kr; hanju.yoo@yonsei.ac.kr; cbchae@yonsei.ac.kr). B. Kim is with SensorView, Ltd., Gyeonggi-do, 13493, South Korea (e-mail: klaus.kim@sensor-view.com). K.-K. Wong is with University College, London, UK, also affiliated with Yonsei Frontier Lab., Yonsei University, Seoul 03722, South Korea (e-mail: kai-kit.wong@ucl.ac.uk).}
\thanks{Manuscript received XXX XX, 2023; revised XXX XX, 2023.}}

%
%

\markboth{}%
{Shell \MakeLowercase{\textit{et al.}}: Bare Demo of IEEEtran.cls for IEEE Communications Society Journals}
%



\maketitle

\begin{abstract}
Full-duplex (FD) technology is gaining popularity for integration into a wide range of wireless networks due to its demonstrated potential in recent studies. In contrast to half-duplex (HD) technology, the implementation of FD in networks necessitates considering inter-node interference (INI) from various network perspectives. When deploying FD technology in networks, several critical factors must be taken into account. These include self-interference (SI) and the requisite SI cancellation (SIC) processes, as well as the selection of multiple user equipment (UE) per time slot. Additionally, inter-node interference (INI), including cross-link interference (CLI) and inter-cell interference (ICI), become crucial issues during concurrent uplink (UL) and downlink (DL) transmission and reception, similar to SI. Since most INI is challenging to eliminate, a comprehensive investigation that covers radio resource control (RRC), medium access control (MAC), and the physical layer (PHY) is essential in the context of FD network design, rather than focusing on individual network layers and types. This paper covers state-of-the-art studies, including protocols and documents from 3GPP for FD, MAC protocol, user scheduling, and CLI handling. The methods are also compared through a network-level system simulation based on 3D ray-tracing.
\end{abstract}

\begin{IEEEkeywords}
Full-duplex (FD), Flexible duplex, Advanced duplex, Network design, Protocol, RRC, PHY, MAC
\end{IEEEkeywords}

%
\IEEEpeerreviewmaketitle

\section{Introduction}
%
%
%
%
Full-duplex (FD) systems, which have been demonstrated to double spectral efficiency through proof-of-concept (PoC) results~\cite{FD6G,FD,FDchae,FDRs}, are currently under consideration for application at the network-level~\cite{3gppDupEn}. The implementation of the FD network, however, requires changes in existing signal processing, protocols, and scheduling to manage cross-link interference (CLI), including self-interference (SI)~\cite{202102wcom, advD, kim2015survey}. The demand of the FD network calls for in-depth discussions in the physical (PHY) layer, medium access control (MAC) layer, and radio resource control (RRC) layer~\cite{FDmac_Mag}.

Wireless networks applying FD technology fall into two main categories: cellular networks, which encompass the 6th generation (6G), and wireless local area networks (WLAN). Cellular or mobile networks are extensive wireless communication systems designed to provide broad area coverage and mobility support for mobile devices. These networks consist of interconnected base stations (BS) or cell sites spread across a vast geographical area, all linked to a core network. Cellular networks utilize licensed frequency bands, including millimeter-wave (mmWave), sub-6GHz, and upper-mid band~\cite{NR}. The evolution of cellular network has been well known as global system for mobile communications (GSM), code division multiple access (CDMA), long term evolution (LTE), 5G, and 6G. WLAN, on the other hand, is a smaller-scale wireless network covering a confined area, such as a home, office, or building. WLANs operate on unlicensed frequency bands and adhere to the IEEE 802.11 standards, commonly known as Wi-Fi~\cite{IEEEwifi}. 

Modern cellular networks must manage diverse types of traffic, focusing more on ultra-reliable low-latency communication (URLLC)~\cite{KKS_PIEEE}. The diversified traffic is described with enhanced mobile broadband (eMBB) and massive machine-type communications (mMTC) in 5G and beyond, necessitating technologies like network slicing~\cite{NR,3gppmag}. {MmWave technology} has brought substantial performance improvements in 5G new radio (NR), but also poses challenges in coverage~\cite{bai2013cov_mmW}. To accommodate diversified traffic, technologies such as reconfigurable intelligent surfaces (RIS) and unmanned aerial vehicle (UAV)-based BS were proposed~\cite{LingRIS, RISNOMA_hwi}.

Beyond doubling spectral efficiency, FD technology can offer flexible network designs~\cite{advD}. Both the 3rd generation partnership project (3GPP) and IEEE 802.11 are actively discussing the use of FD in cellular networks and Wi-Fi networks, respectively~\cite{3gppDupEn,XDD,202102wcom}. The evolution of complex nonlinear signal processing technology resulted in self-interference cancellation (SIC) performance of over $110$~dB, suggesting the feasibility of operating FD in a real BS~\cite{NLSI,NLSIsim,202102wcom}.

{The attention is shifting towards FD network design, which spans across multiple BS and extends beyond the limited scope of the PHY layer to encompass the MAC and RRC layers~\cite{heteroFD,flex1,op}. This network design represents the next phase after dealing with signal processing in the PHY layer, enhancing radio frequency (RF) hardware components, and validating PoC. Initially, the focus was primarily on assessing FD technology's performance and feasibility at the individual device level or within constrained scenarios. However, once the foundational challenges and limitations have been addressed on the link-level, the transition to network design becomes imperative to fully unlock FD's potential in large-scale, real-world deployments.} This article aims to introduce and provide insight into FD-enabled network design, which covers multiple BS and multiple access.

\begin{figure*}

	\begin{center}
		{\includegraphics[width=2\columnwidth,keepaspectratio]
			{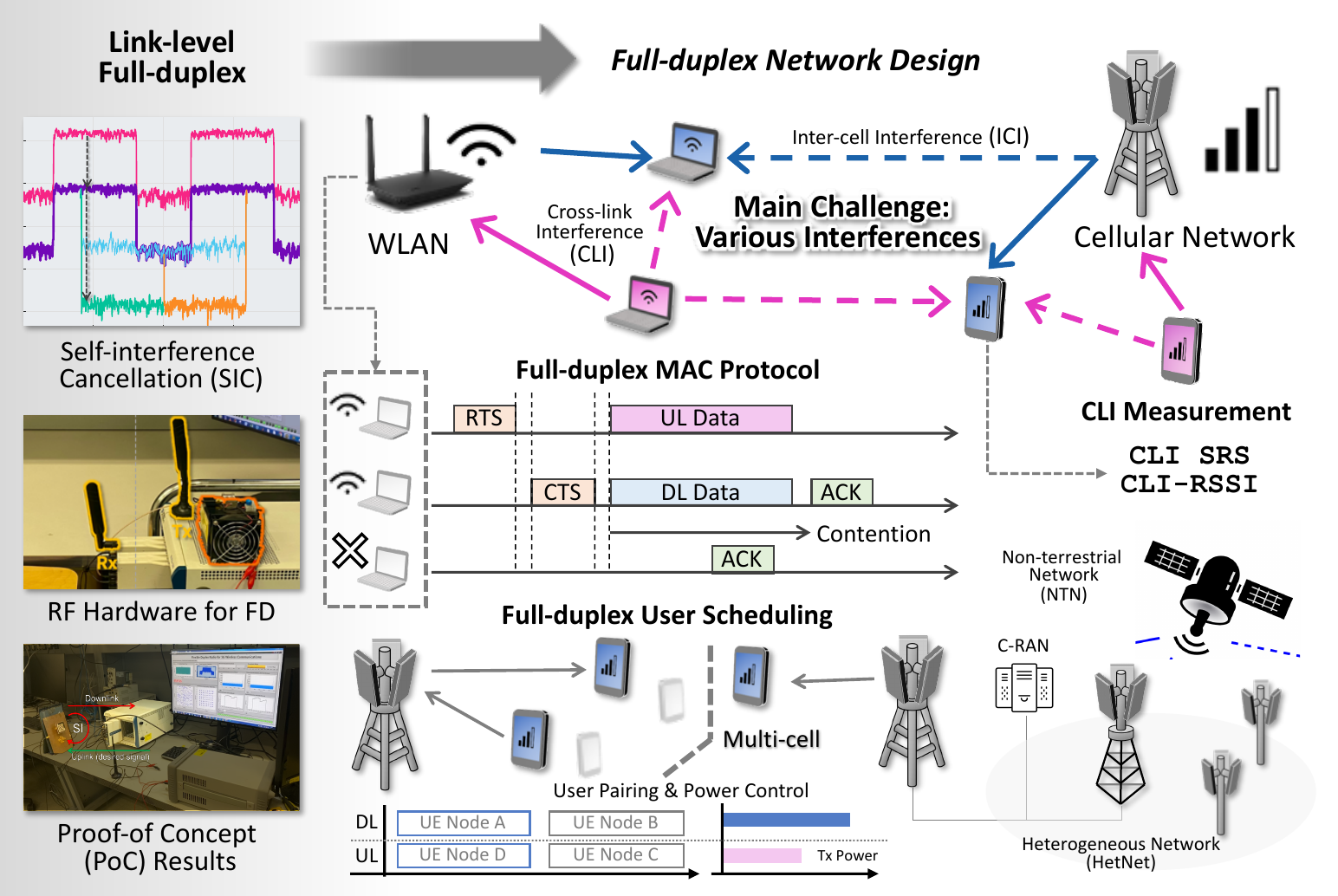}%
			\caption{Challenges for full-duplex network design.}
			\label{fig.challenges}
		}
	\end{center}

\end{figure*}

\subsection{Challenges of Full-duplex Network Design}

FD network design has maximized its necessity and timeliness through successful research results, such as signal processing including SIC, antenna theory and process, and practical performance verification through PoC. There are challenges in FD network design that can be identified through state-of-the-art works. The following challenges are the implementation requirements for each node when incorporating FD into a real network:
\begin{itemize}
\item {MAC protocol: Existing half-duplex (HD) protocols are already complex. In FD, the complexity becomes even more pronounced because the number of links involved in a single FD communication scenario can vary, even when there is the same number of nodes. Various studies, including verification through PoC in selected scenarios, have been conducted.}
\item {SIC for various FD scenarios: SIC usually operates at the BS node with a relatively large transmit power, leading to nonlinearity. Ensuring stable analog and digital SIC is essential for various FD duplex modes, including in-band FD (IBFD) and sub-band FD (SBFD).}
\item Fairness control: In contrast to link-level communication, where the goal is to maximize the data-rate, actual networks require consideration of fairness for multiple users. Fairness in FD networks presents a more complex issue than in HD networks, as it requires simultaneous consideration of both uplink (UL) and downlink (DL).
\end{itemize}

{Fig.~\ref{fig.challenges} illustrates that the primary challenge in FD networks is addressing inter-node interference (INI). While INI was present in HD networks due to the multi-cell environment, it becomes more significant in FD networks because of simultaneous transmission and reception. Managing INI at each node is a complex task, unlike SI, and this gives rise to the following factors to consider in FD network design:}

\begin{itemize}
\item Cross-link interference (CLI) management: {In a multi-user FD scenario, CLI can result in DL interference, potentially degrading the overall system throughput. Therefore, in an FD system, it is important to choose the appropriate user equipment (UE) based on the network topology and the resulting CLI. This adds complexity to scheduling when compared to HD scenarios, making the search for optimal scheduling a significant challenge.}
\item Inter-cell interference (ICI) issue in multi-cell: Practical cellular networks consist of multi-cell structures, and any cell and BS node within it can have a considerable influence on the FD operation of neighboring cells. {To address the ICI issue, it is essential to extend the current solutions used in HD scenarios, such as coordinating between BS nodes, to fulfill the requirements in FD scenarios.}
\item {Directivity of INI and SI: A multiple-input multiple-output (MIMO) beamforming system introduces beam directivity. These directional beamforming techniques in FD systems can be used to minimize both INI and SI, enabling a potential low-complexity solution.} 
\end{itemize}

Through the remaining parts of this section, we will first examine the existing literature that comprehensively addresses the above issues, and then explain the direction of this article.

\subsection{Related Studies on Full-duplex Network Design}
{Optimizing the PHY layer for FD had been the subject of considerable research across all network layers~\cite{kim2015survey}. For cellular networks, \cite{FDcellular,202102wcom} discussed the impact of INI on cellular networks in the network level and compared the overall network design for cellular networks to HD communication. Duplex enhancement, which includes FD communication, is also investigated by various industries~\cite{XDD,simQ} and is currently considered a work item of the 3GPP release 18~\cite{3gppDupEn}.}

In the case of WLAN, the primary focus of FD network design had been on the MAC protocol. User scheduling in WLAN operates through competition among nodes in the network based on carrier sensing. These protocol design for FD network is studied rapidly by many researchers~\cite{FDmac_Mag,FDmac_wn}. Active research had been carried out to implement FD communication with minimal changes to the existing HD protocol~\cite{kim2017asymmetric}. 

{There is also pioneering work that has addressed routing protocols in the context of multi-hop FD networks~\cite{david2013optimal, moein2018bidrectional, kadri2019xfdr}. 
In multi-hop scenarios with FD capable links from source to destination nodes, routing involves considerations distinct from traditional HD-based protocols, requiring collaboration across both MAC and PHY layers. FD's advantages in routing include immediate forwarding, continuous sensing, and faster convergence. FD technology's ability to enable simultaneous transmission and reception is highly beneficial in multi-hop wireless networks, allowing for rapid data forwarding. FD also allows for medium sensing while transmitting, enhancing network responsiveness and adaptability. The combined effect of immediate forwarding and continuous sensing accelerates the dissemination of routing protocol signaling, leading to quicker network convergence.}

 {From an algorithmic standpoint, routing in FD networks necessitates adaptations to traditional routing algorithms like Dijkstra's or Bellman-Ford. The dynamic nature of SI and CLI, along with fluctuating link quality depending on transmission direction, calls for centralized control mechanisms that can rapidly adapt. Routing challenges in FD become prominent in networks where FD offers significant advantages, such as integrated access and backhaul (IAB) systems and heterogeneous networks (HetNets). These areas present fertile ground for future research and development.
}

\subsection{Contributions and Organization of This Article}

 {In this article, we aim to address three major points concerning FD network design. A common goal is to manage interference between FD network nodes, INI. The conventional survey studies focus on specific network types and layers. It is timely to cover state-of-the-art research to provide a comprehensive understanding of FD network design. A typical network spans from RRC to MAC and PHY layer and can be divided into WLAN and cellular network categories based on configuration. We begin by introducing the recent FD MAC protocol in WLAN, followed by FD scheduling in cellular network from an algorithmic perspective, and cover CLI handling issues, depending on the network structure. We finally examine the comprehensive network-level evaluation beyond link-level FD.} The contributions of this article include:

\begin{itemize}
\item First, we discuss the state-of-the-art MAC protocol for FD networks, especially for WLAN. Research into FD MAC protocols must strive for compatibility with existing HD protocols and standards, while also taking into account various forms of INI. In this context, we present recent advancements in FD MAC studies, focusing on how they establish and synchronize node connections. 
\item Second, we address the user scheduling issue of the cellular network. 
The user scheduling issues, in particular, require centralized control to manage INI, and concepts such as network-assisted FD (NAFD) have been proposed. The optimization problem is a non-convex issue necessitating a computational theory approach, such as nondeterministic polynomial (NP), from an algorithmic perspective. 
\item Third, we present the special network concept and strategies to tackle the CLI handling issue. CLI is challenging to measure and cancel, unlike SI, and occurs much more in FD networks than in conventional HD networks. 
To address this issue, various network structures reliant on centralized control and coordination among multiple nodes have been proposed. The novel structures and RRC signaling are notably suitable for FD operation, and its potential has been verified in numerous studies.
\item Lastly, we provide the 3D ray-tracing-based system-level simulation (SLS) that aware the multi-cell multi user FD network. We consider the analog beamforming on spatial domain examined by real-world channel. The impact of the analog beamforming emerges in INI handling challenge. We expect that the comprehensive SLS can provide the insight for FD network design.
\end{itemize}

This article is organized as follows: After introducing general FD basics and other signal processing technologies to implement FD in Section~\ref{sec.basic}, we first provide the MAC perspective to establish the FD link between nodes in Section~\ref{sec.MAC}. In Section~\ref{sec.US}, we review the user scheduling researches of FD network design. We accordingly present the CLI handling issue that is essential for both FD network protocol and user scheduling in Section~\ref{sec.CLI}. In Section~\ref{sec.EVAL}, we evaluate the system level, network performance of state-of-the art methods through 3D ray-tracing-based channel realization, and we conclude this article through Section~\ref{sec.CNCL}.

\begin{figure*}[t]
	\begin{center}
		\subfigure[WLAN full-duplex system.]{\includegraphics[width=0.7\columnwidth,keepaspectratio]
			{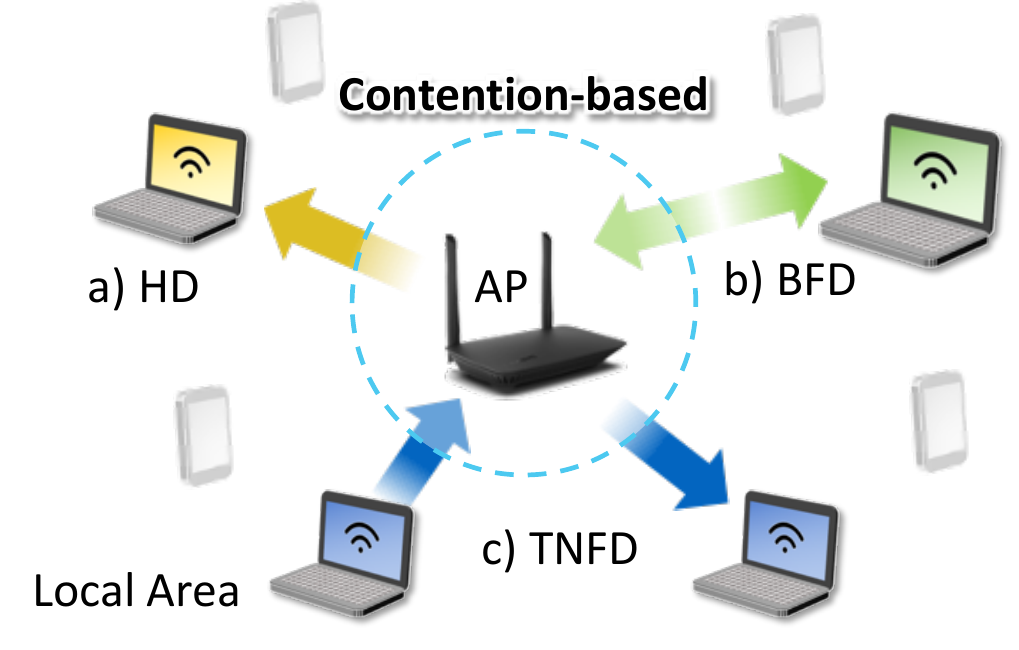}%
			\label{fig.WLAN}
			}
		\subfigure[Cellular network full-duplex system.]{\includegraphics[width=1.3\columnwidth,keepaspectratio]
			{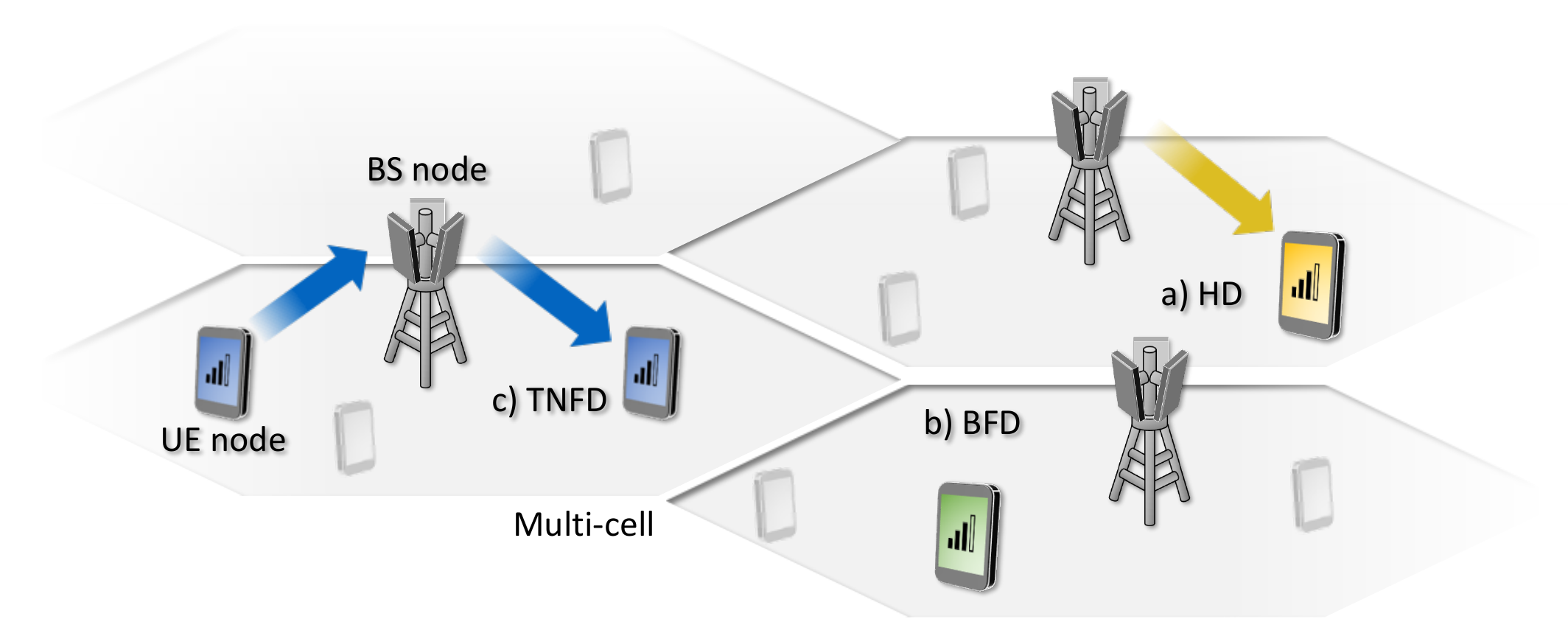}%
			}
		\subfigure[Half duplex, downlink (DL) mode.]{\includegraphics[width=.62\columnwidth,keepaspectratio]
			{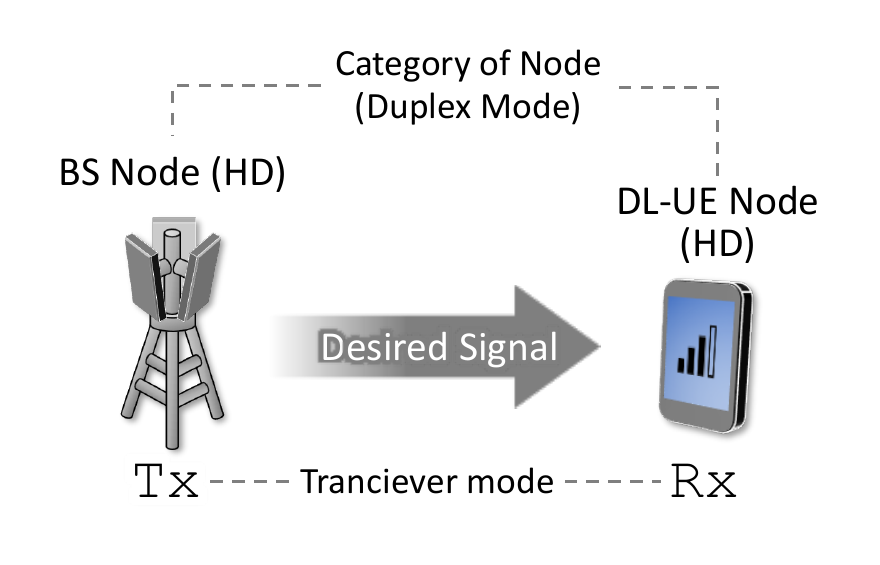}%
			\label{Fig.HDmode}
			}
		\subfigure[Symmetric FD or bi-directional FD (BFD).]{\includegraphics[width=.62\columnwidth,keepaspectratio]
			{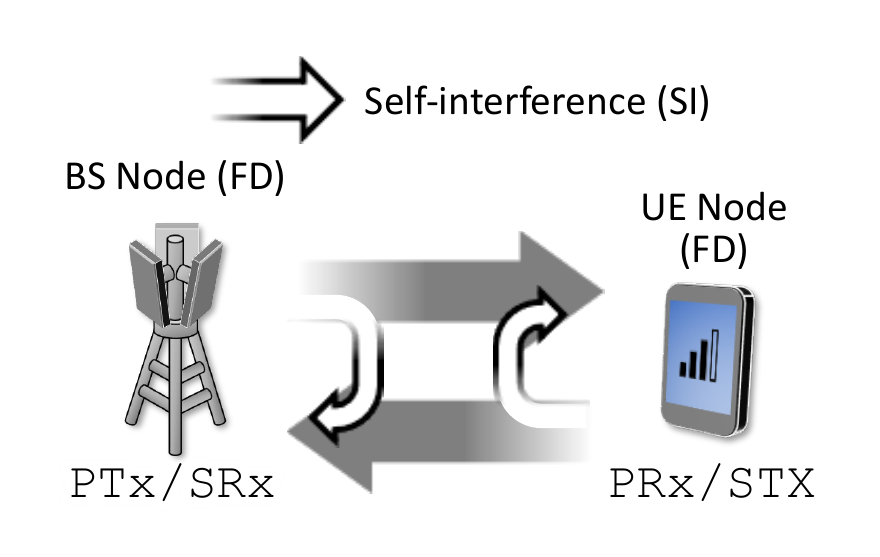}%
			\label{fig.BFD}
			}
		\subfigure[Asymmetric FD or three node FD (TNFD).]{\includegraphics[width=.74\columnwidth,keepaspectratio]
			{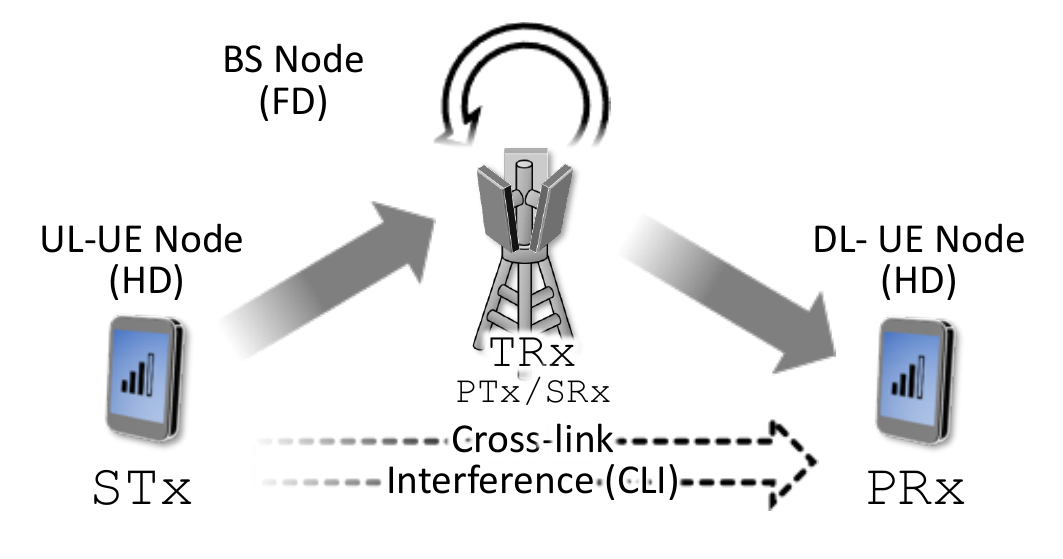}%
			\label{fig.TNFD}
			}
		\subfigure[Source-based transmission mode (SBTM) for TNFD.]{\includegraphics[width=\columnwidth,keepaspectratio]
			{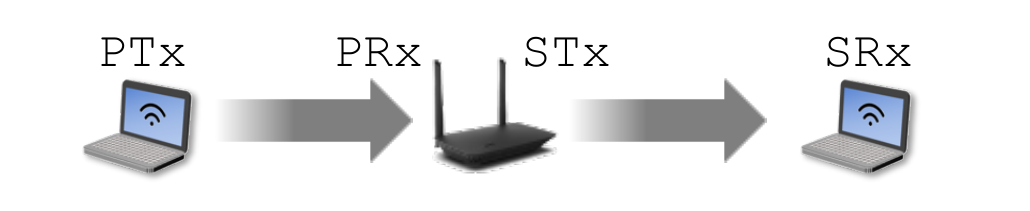}%
			\label{fig.SBTM}
			}
		\subfigure[Destination-based transmission mode (DBTM) for TNFD.]{\includegraphics[width=\columnwidth,keepaspectratio]
			{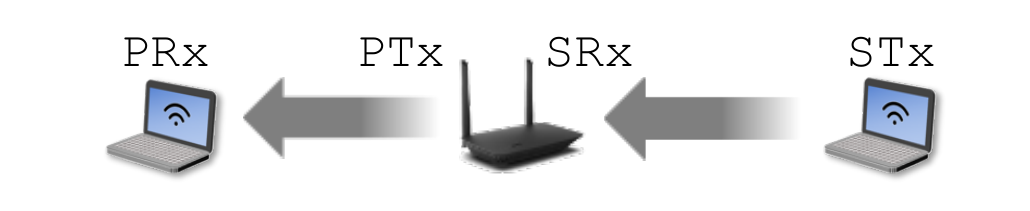}%
			\label{fig.DBTM}
			}
	\end{center}
	\caption{Full-duplex network topology and description of duplex mode.}
\label{fig.nodes}	
\end{figure*}

\subsection{Notations and Abbreviations}

The mathematical notations in this article are as follows. The lowercase, bold lowercase, bold uppercase letters, such as a, $\mathbf{a}$, and $\mathbf{A}$, denote the scalar, vector, and matrix, respectively. $\mathbb{A}$ denotes a set. Also, since duplex is dealt with in this article, the type of link is expressed as the direction of signal transmission. Unless otherwise stated, the superscript of a specific variable indicates the classification of that variable, and the subscript indicates the index. $a\rightarrow b$ is $a$ sending a signal to $b$, and $a\leftarrow b$ is $a$ receiving a signal from $b$. For example, $p^{\text{BS}\leftarrow \text{UE}}_{a\leftarrow b}$ is the UL signal and transmit power on the link from UE node $b$ to BS node $a$. $\mathbb{R}$ and $\mathbb{C}$ denote the real and complex number sets, respectively, and $\mathbb{C}^{M\times N}$ denotes the space of $M\times N$ complex-valued matrices. $j\triangleq\sqrt{-1}$ is the imaginary unit. 

All nodes, including BS, access point (AP), and UE nodes, can be designated as either a transmitter ($\mathtt{Tx}$), a receiver ($\mathtt{Rx}$), or both (i.e., transceiver, $\mathtt{TRx}$). To express both classifications simultaneously, we use the term ``duplex mode-node role (transceiver mode)", such as FD-BS node ($\mathtt{TRx}$). HD mode UEs matched to an FD-BS node are classified into UL-UE node and DL-UE node for each UL/DL, respectively. In addition to the $\mathtt{Tx}$ and $\mathtt{Rx}$ classification, the FD protocol designates primary and secondary nodes, resulting in a total of four node classifications. Primary $\mathtt{Tx}$ ($\mathtt{PTx}$) and primary $\mathtt{Rx}$ ($\mathtt{PRx}$) initiate transmission and reception first, followed by secondary $\mathtt{Tx}$ ($\mathtt{STx}$) and secondary $\mathtt{Rx}$ ($\mathtt{SRx}$). In FD scenarios and general HD scenarios that do not consider protocols, each node is divided into FD and HD nodes without distinguishing between primary and secondary, and separated by $\mathtt{Tx}$ and $\mathtt{Rx}$. A typical FD pair consists of UL-UE node ($\mathtt{Tx}$) $\rightarrow$ (FD)-BS node ($\mathtt{TRx}$) $\rightarrow$ DL-UE node ($\mathtt{Rx}$).

\section{Basics of Full-duplex:\\Network Deployment and SIC}\label{sec.basic}

\subsection{Network Deployment and Components}

In network design, a multi-user scenario is assumed with the goal of applying it to actual cellular networks. 
FD communication in a single wireless channel has the potential to double the spectral efficiency compared to HD systems. Legacy systems, such as time-division duplex (TDD) and frequency-division duplexing (FDD), separate signals either by time-switching or filters, each having their advantages and disadvantages. 
Advanced duplex systems, such as cross-division duplex~\cite{XDD} and flexible duplex~\cite{alpha1,US-alpha}, had been proposed to address these challenges.

The nodes that make up the FD network can be classified into three categories: node role, duplex mode, and transceiver mode, as shown in Fig.~\ref{fig.nodes}. From the viewpoint of node role, nodes can be a BS, UE, or AP. For duplex mode, BS nodes are usually compatible with FD mode, while UE nodes are only compatible with HD mode. More specifically, there are different FD configurations, such as symmetric FD or bi-directional FD (BFD) and asymmetric FD or three-node FD (TNFD). 
TNFD can be divided into source-based transmission mode (SBTM) and destination-based transmission mode (DBTM) from the MAC point of view.

Fig.~\ref{fig.BFD} illustrates the BFD mode, where a reference BS node exchanges UL/DL signals with a single UE node. {In BFD, managing interference between both nodes establishing a link becomes more straightforward when both nodes have SIC capabilities, as BFD introduces only SI and does not involve CLI.} However, considering the complexity of implementing SIC in a UE node, such as a mobile device, and the high probability of asymmetric traffic requirements across nodes from a network perspective, the likelihood of forming BFD is lower than that of TNFD.

Fig.~\ref{fig.TNFD} represents the TNFD mode, in which three nodes participate. Both the UL-UE node transmitting the UL signal and the DL-UE node receiving the DL signal coexist in the communication. As shown in Fig.~\ref{fig.TNFD}, the signal transmitted from UL-UE causes CLI in DL-UE. We can categorize TNFD into SBTM and DBTM from the MAC perspective. 

Fig.~\ref{fig.SBTM} depicts SBTM, which establishes a link based on the node with traffic to transmit, a kind of source. We set the UL-UE node to $\mathtt{PTx}$ for primary transmission to synchronize first, designate the BS node as $\mathtt{PRx}$ and $\mathtt{STx}$, and assign the DL-UE node as $\mathtt{SRx}$. Fig.~\ref{fig.DBTM} shows DBTM, which sets the DL-UE node as $\mathtt{PRx}$ and the UL-UE node as $\mathtt{STx}$. We will discuss the MAC classification aspects of TNFD, the process of assigning nodes, and link formation in more detail in Section~\ref{sec.MAC}.


\subsection{ {Link-level Full-duplex System and Self-interference}}
 {
Each BS operates an orthogonal frequency division multiplexing (OFDM) system with subcarriers, which enable orthogonal channel realizations. The single time slot of each subcarrier is also referred to as resource element (RE), and the set of RE is resource block (RB) in OFDM standard. 
SIC mainly intervenes on the BS node. An arbitrary BS node (FD, $\mathtt{TRx}$), $b$, simultaneously transmits a DL signal to a DL-UE node, $i$, and receives a UL signal, $y^\text{UL}_{b\leftarrow j}$, from a UL-UE node, $j$. SI at $b$, $y^\text{SI}$, however, interferes UL signal, so the received signal at BS node $b$, $y^\mathtt{Rx}_{b\leftarrow j}$ is as follows:
\begin{equation}
\begin{aligned}
y^\mathtt{Rx}_{b\leftarrow j}=y^\text{UL}_{b\leftarrow j}+y^\text{SI}_b+n_b,
\end{aligned}
\end{equation}
where $n_b$ is noise at the BS node receiver. 
The channel gain between BS node $b$ and UE node $i$ is given as, $G^\text{BU}_{b,i}$.
At this point, the gain of the SI signal at the BS node $b$, $G^\text{SI}_b$, includes both analog and digital SIC. The BS node transmits the signal, $x^\text{DL}_{b\rightarrow i}$, with power $p^\text{DL}_b$ and this component generates the SI as follows:
\begin{equation}
y^\text{SI}_b=\sqrt{G^\text{SI}_b p^\text{DL}_b }x^\text{DL}_{b\rightarrow i}
\end{equation}
SIC performance, $G^\text{SI}_b$, is the main purpose of the link-level FD studies~\cite{NLSI}. The total power of residual SI, $I^\text{SI}_\text{BS}$ for FD-compatible BS node $b$ is as follows:
\begin{equation}
I^\text{SI}_\text{BS}=G^\text{SI}_b p^\text{DL}_b.
\end{equation}
Antenna and RF component structures play a crucial role in enabling FD communication~\cite{NLSIsim,202102wcom}. The isolation between the $\mathtt{Tx}$ and $\mathtt{Rx}$ is critical for mitigating SI, which can be achieved through antenna separation or active cancellation techniques. Researchers have demonstrated considerable cancellation performance through multiple or single antennas in various PoC results, including SI due to the power amplifier (PA) nonlinearity~\cite{XDD}.
}
\subsection{Inter-node Interferences on Full-duplex Network}

INI can be broadly categorized into CLI between UE nodes, and ICI between BS nodes. CLI and ICI, unlike SI, are difficult to identify and remove, and can significantly impact the signal-plus-interference-to-noise ratio (SINR). The FD network design, thus, requires INI management from a network perspective. 
The overall multi-cell network is defined as follows: The network consists of BS node set $\mathbb{B}$, each BS node $b\in\mathbb{B}$ represents each cell, and BS node $b$ services the set of UE nodes, $\mathbb{U}_b$, for both UL and DL. 

CLI primarily occurs in asymmetric FD situations. An example of CLI is when a signal originating from a UL-UE node interferes with a DL-UE node in an inter-cell or shared AP situation. The BS node $b$ schedules UE nodes of both UL and DL as $\pi_b$, by associating UL-UE node $j=\pi_b(i)$ to each DL-UE node, $i$. The channel gain between UE nodes $i$ and $j$, $G^\text{UU}_{i,j}$, then generates CLI as $I^\text{CLI}_{\text{UE}\leftarrow\text{UE}}$. 

ICI occurs in a multi-cell environment, even in the existing HD cellular network system. A DL signal transmitted from a BS node $b'$ of another cell may act as interference to the current BS node $b$, as $I^\text{ICI}_{\text{BS}\leftarrow\text{BS}}$. The same signal from BS node $b'\in\mathbb{B}$ acts as an ICI to a UE node $j$ as $I^\text{ICI}_{\text{UE}\leftarrow\text{BS}}$. ICI in FD is similar to HD but happens more intensely. Handling interference becomes trickier when both the BS node and UE nodes transmit and receive simultaneously because both types of ICI can occur at the same time.

The effect of INI, including SI, CLI, and ICI, on the received SINR of each node can be examined as follows. We define the strength of the desired signal of UL and DL as $S^\text{UL}$ and $S^\text{DL}$, respectively. First, the SINR for DL-UE node $i$, which is serviced from BS node $b$ and paired with UL-UE node $j$, is denoted as follows: 
\begin{equation}
\begin{aligned}
\text{SINR}_{b\rightarrow i}^\text{DL} 
&=\frac{S^\text{DL}_{\text{BS}\rightarrow \text{UE}}}{I^\text{CLI}_{\text{UE}\leftarrow \text{UE}}+I^\text{ICI}_{\text{UE}\leftarrow\text{BS}}+\mathcal{N}_\text{UE}}\\
&= \frac{p_{b}^\text{DL}G_{b,i}^\text{BU}}
{\underbrace{p_{j}^\text{UL}G_{i,j}^\text{UU}}_{\text{CLI: UE-to-UE}}+\underbrace{\displaystyle\sum_{b'\in\mathbb{B}}p_{b'}^\text{DL}G_{b',i}^\text{BU}}_{\text{ICI: BS-to-UE}}+
\mathcal{N}_\text{UE}}.
\end{aligned}
\end{equation}
The UL SINR for UE node $j$, at the same time, is expressed as follows:

\begin{equation}
\begin{aligned}
\text{SINR}_{b\leftarrow j}^\text{UL} 
&=\frac{S^\text{UL}_{\text{BS}\leftarrow\text{UE}}}{I^\text{ICI}_{\text{BS}\leftarrow\text{BS}}+I^\text{CLI}_{\text{BS}\leftarrow\text{UE}}+I^\text{SI}_\text{BS}+\mathcal{N}_\text{BS}}\\
&= \frac{p_{j}^\text{UL}G_{b,j}^\text{BU}}
{
\underbrace{\displaystyle\sum_{b'\in\mathbb{B}}p_{b'}^\text{DL}G_{b',j}^\text{BU}}_{\text{ICI: BS-to-BS}}+\underbrace{p_b^\text{DL}G^\text{SI}_b}_{\text{SI (residual)}}+\mathcal{N}_\text{BS}}.
\end{aligned}
\end{equation}
From the SINR, we can calculate the UL/DL throughput, $R_j^\text{UL}$ and $R_i^\text{DL}$, as follows:
\begin{equation}
\begin{aligned}
R_j^\text{UL} = \text{BW}_j^\text{UL}\log(1+\text{SINR}^\text{UL}_{b\leftarrow j}), \\
R_i^\text{DL} = \text{BW}_i^\text{DL}\log(1+\text{SINR}^\text{DL}_{b\rightarrow i}).
\end{aligned}
\end{equation}
Various duplex modes can be realized by controlling the UL/DL bandwidth (BW) allocation ratio. Duplex modes include FD-mode, HD-mode, and flexible duplex.




\subsection{Proof-of-Concept Results on Full-duplex}

{The successful validation of FD's ability to nearly double spectral efficiency compared to HD through PoC has greatly enhanced its potential integration into wireless networks. PoC studies have explored the applicability of FD in various network types, including cellular networks and WLANs, highlighting its suitability for these contexts.}

 {Implementation through PoC requires the incorporation of digital and analog SIC technology as well as protocol implementation in the MAC and PHY layers. In \cite{FDRs}, the first FD implementation paper, a basic single-carrier IBFD system was built based on the IEEE 802.11ac standard. This study, which assumed a single-input-single-output (SISO) antenna, implemented a circuit board and a prototype compatible with the existing Wi-Fi system. Additionally, in \cite{sahai2011pushing}, an FD PoC was implemented as an OFDM multi-carrier system, presenting a MAC protocol suitable for IEEE 802.11 with header snooping and other features. Overall, FD PoC demonstrated a 1.7-fold improvement in spectral efficiency.}

 {The authors in \cite{FDchae} demonstrated the spectral efficiency of FD improving to a specific value of $1.9$~times, focusing on SIC and targeting 5G cellular networks. Using a software-defined radio (SDR) platform, the researchers built a prototype equipped with both analog SIC through a dual-polarization antenna and real-time digital SIC. Furthermore, they extended the antenna configuration to MIMO through \cite{compactFDd2d} and verified the FD performance in a device-to-device (D2D) scenario using a field-programmable gate array (FPGA)-based SDR. Studies are underway to confirm the effects of SI and SIC in mmWave FD communications through prototyping such as mmWave~\cite{IABgy,simQ,XDD}.}


\subsection{Current Standard for Full-duplex Network}

In terms of the latest cellular network standards, the FD network standard for duplex enhancement is being discussed as a work item in 3GPP Release 18 from 2022, with details actively being updated. In the NR standard documents, three FD network scenarios are presented. {These options include choosing a uniform duplex mode, whether FD or HD, for all BS nodes (cells), or integrating both FD and HD BSs.} In the documents, interference between nodes is generally referred to as CLI. The types of interference are: 1) BS (gNB) Self-interference, 2) Inter-cell BS-BS inter-subband CLI, and 3) UE-UE co-channel inter-subband CLI.

For BS-to-BS and UE-to-UE channels, large-scale fading must be modeled, and for BS-to-BS CLI, antenna gain and multi-path channel modeling must be followed. The performance metric of the duplex enhancement system includes throughput, measured as the mean or median value of user perceived throughput (UPT), and latency, evaluated as packet-latency or UE-average-latency. Packet latency is defined as the time starting when the packet is received in the transmit buffer and ending when the last bit of the packet is correctly delivered to the receiver. Section~\ref{sec.CLI} covers potential enhancements of dynamic TDD and flexible TDD, comparing them with FD.

Nonlinear SI due to hardware impairments in $\mathtt{Tx}$ is defined as adjacent-channel CLI and handled by digital SIC. Each subcarrier in OFDM causes interference to adjacent subcarriers, and several early-stage studies aimed to address this issue~\cite{K_iter,FDC_iter}. The FD node employs a structure that separates $\mathtt{Tx}$ and $\mathtt{Rx}$. Additionally, by partitioning all antenna elements, a transceiver structure is designed by assigning half of $\mathtt{Tx}$ for DL in the FD node or BS and allocating $\mathtt{Rx}$ for UL.

The existing 3GPP standard for duplex scheduling appeared in \cite{3gppSche}, which had been in progress since 2017. To efficiently allocate a given frequency-time resource, including UL and DL, dynamic scheduling is performed in the MAC of the BS node. The scheduler allocates radio resources based on the buffer status and quality of service (QoS) requirements of each UE node or through the radio condition measured by the BS node and each UE node.

In the case of DL scheduling, the BS node dynamically allocates resources to the UE node through the radio network temporary identifier (RNTI) of the physical DL control channel (PDCCH). For UL, PDCCH is also monitored to check if the resource is available for UL transmission. The CLI handling issue in the existing duplex modes is also covered in the standards. Unlike FD, in the case of HD, only CLI generated from neighboring cells or BS-nodes can be experienced according to various TDD patterns. To mitigate CLI, BS nodes can send and receive TDD patterns, that is, UL and DL arrangements with each other, through the Xn and F1 interfaces. The victim UE can be designated and used for CLI estimation, with two estimation methods: SRS-RSRP and CLI-RSSI, we will discuss the details in Section~\ref{sec.CLI}.

Furthermore, through \cite{BWP1,BWP2}, 5G NR supports various numerologies for subcarriers even in the same time slot. This indicates that the $\mathtt{Tx}$ operation and the $\mathtt{Rx}$ operation can be performed differently or simultaneously, depending on the frequency or subcarrier for a specific node, which can be the one of the application of FD to the network.

\section{Scheduling Designs for \\Full-duplex WLAN}\label{sec.MAC}


\subsection{ {Challenges of Full-Duplex MAC Protocols}}

Traditional MAC algorithms were designed for HD communication, where nodes could not transmit and receive simultaneously on the same channel. However, as the nodes can transmit and receive simultaneously on the same channel, FD communication introduces several challenges, such as inter-node interference, hidden node problem in FD environments, and fairness between HD and FD nodes~\cite{FDmac_Mag}. These challenges could not be efficiently handled by conventional HD MAC protocols like carrier-sensing multiple access with collision avoidance (CSMA/CA)~\cite{kim2015survey}. Therefore, the need for an efficient MAC protocol for FD communication has arisen to address these challenges and to fully realize the benefits of FD communication. In the following paragraph, we introduce the new challenges that the emergence of FD links has brought to medium access control.

\begin{figure*}
\begin{center}
    \subfigure[Inter-node interference.]{%
      \includegraphics[height=0.55\columnwidth,keepaspectratio]
      {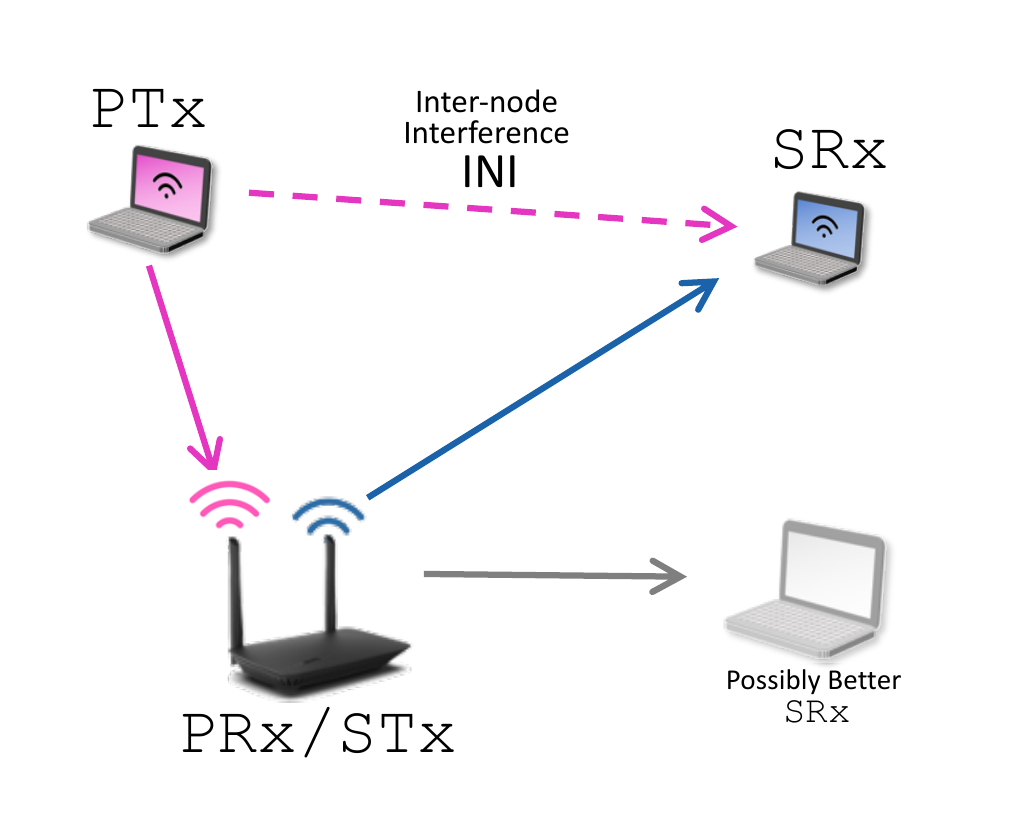}%
      \label{subfig:inter_node_interference}
    }
    \subfigure[Hidden node problem.]{%
      \includegraphics[height=0.55\columnwidth,keepaspectratio]
      {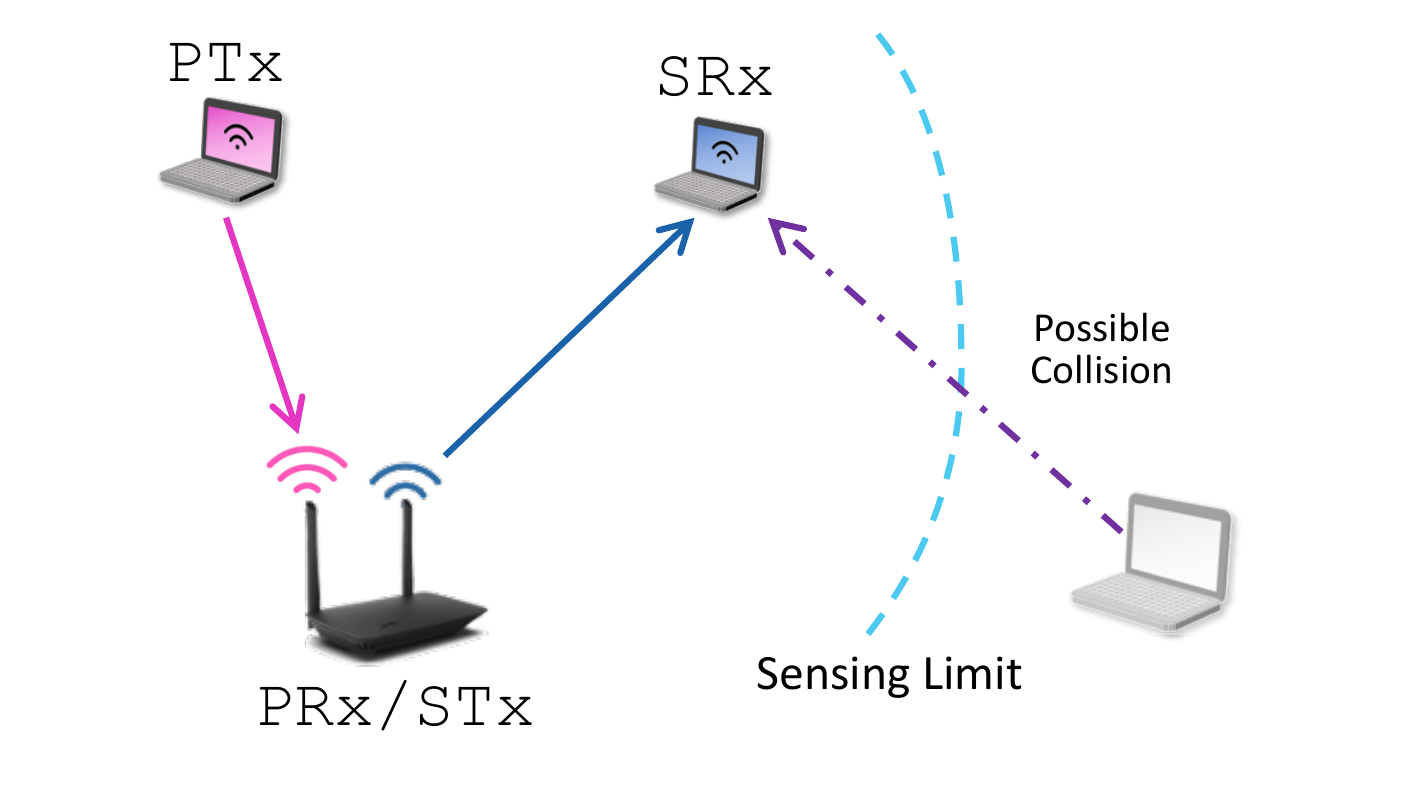}%
      \label{subfig:hidden_node}
    }
 \end{center}
	\caption{Illustration of inter-node interference and hidden node problem in full-duplex communications.}
	\label{fig:fd_link_types}
\end{figure*}

\subsubsection{Inter-node interference (INI)}

Inter-node interference (INI) poses a significant challenge in FD communication networks. It occurs when the $\mathtt{PTx}$ signal interferes with the $\mathtt{STx}$ signal, leading to a collision in a $\mathtt{SRx}$ in TNFD scenarios, as described in Fig.~\ref{subfig:inter_node_interference}. Due to the inability of the $\mathtt{SRx}$ to distinguish between the $\mathtt{PRx}$ signal and the $\mathtt{STx}$ signal, the SINR may degrade, leading to a reduced data rate or spectral efficiency~\cite{xie2014does}. Hence, selecting the $\mathtt{SRx}$ carefully is crucial to minimize INI and maximize overall spectral efficiency.

To select good candidates for the $\mathtt{SRx}$, the $\mathtt{STx}$ may use a signal-to-interference-ratio (SIR) map to compute the expected SINR of the transmitted signal at candidate receivers, and then select the one that maximizes the sum spectral efficiency~\cite{goyal2013distributed, tang2015duplex, kim2017asymmetric, liu2019hidden, kim2020performance}. To collect this information, the FD MAC design should consider how to compute SIR information with minimal control signal overhead or computational burden to any single node. The design should also consider how to exchange the SIR packets between nodes, since every node can be a possible $\mathtt{STx}$ and needs to maintain this information for all possible nodes. 

Another possible way to mitigate INI is to compute an optimal transmit power that is sufficient to convey packets to their receivers error-free at the required rate, while still low enough to avoid interference with neighboring nodes~\cite{choi2015power, chen2017probabilistic}. The optimal transmit power can be determined by maximizing the overall sum spectral efficiency, while meeting minimum SINR constraints. This optimization problem can be solved using either a numerical interference model or heuristics to reduce computation time. These computations are typically done utilizing the aforementioned SIR map and can be performed in a centralized or distributed manner, depending on the system requirements and constraints.

\subsubsection{Hidden node problem in full-duplex communications}
In wireless networks, hidden node problem refers to the situation where two nodes are unable to sense each other and both initiate transmission at the same time, causing a collision. This problem can be particularly challenging in FD communications where a node may be transmitting and receiving simultaneously, making it difficult for other nodes to detect their presence~\cite{liu2019hidden}. For example, in a WLAN communications scenario where a station (STA) is transmitting to an AP, but another STA is simultaneously transmitting to the AP as in Fig.~\ref{subfig:hidden_node}, the two interfering STAs may be unable to detect each other, causing a collision. This problem becomes more pronounced in FD communication systems, where types of communications links can vary, e.g., BFD or TNFD, and nodes can simultaneously transmit and receive data.

For example, in a BFD situation, the hidden node problem can be efficiently mitigated as the two nodes involved in communication transmit concurrently, blocking any surrounding nodes from initiating another transmission~\cite{choi2010achieving, doost2015performance}. However, since the transmission length of the two concurrent transmitting nodes can vary, the signal length should be equalized. This can be achieved using a busy tone or by broadcasting an request-to-send (RTS) or clear-to-send (CTS) control signal to prevent possible collisions when one transmitter is idle but the other one is still transmitting. This ensures that the channel remains idle until both nodes have completed their transmissions, preventing any collisions due to hidden nodes.

However, the hidden node problem can still occur in a TNFD, just like in HD communications. In TNFD, three nodes are present in the same network, where the middle node only uses its FD capability to relay a HD link from the first node to another. In this scenario, the neighbor nodes of the receiving nodes may fail to sense transmission of the relaying node and start new transmission, resulting in a collision. To mitigate this problem, a similar solution to HD networks can be employed, where collision detection mechanisms such as RTS/CTS control signals can be used to prevent collisions in the end nodes~\cite{sahai2011pushing, duarte2013design}. Therefore, the MAC algorithm for FD networks should be able to detect the type of communication link (BFD or TNFD) and use the appropriate strategy to alleviate the hidden node problem and ensure higher FD communication opportunities.

\begin{table*}[htbp]
\footnotesize{
\caption{Classification of Various Full-Duplex Medium Access Control Designs}
\label{tab:classification_fdmac}
\centering
\begin{tabularx}{\textwidth}{>{\hsize=0.6\hsize}X|>{\hsize=0.7\hsize}X|>{\hsize=0.7\hsize}X|>{\hsize=1.4\hsize}X|>{\hsize=0.7\hsize}X|>{\hsize=1.1\hsize}X|>{\hsize=1.0\hsize}X}
\hline
\textbf{Ref.} & \textbf{Node with FD capability} & \textbf{Transmission Modes} & \textbf{Channel Access} & \textbf{ {Coordination}} & \textbf{Hidden Nodes Mitigation} & \textbf{INI Consideration}\\
\hline
FD-MAC\cite{sahai2011pushing} & All nodes & BFD, SBTM & Contention-based &  {Decentralized} & ACK/CTS (in header) & Yes (By network topology estimation) \\
\hline
ContraFlow\cite{singh2011efficient} & All nodes & BFD, DBTM & Contention-based &  {Decentralized} & Busy tone & Yes (By $\mathtt{STx}$ rule) \\
\hline
\cite{duarte2013design} & All nodes & BFD & Contention-based &  {Decentralized} & RTS/CTS & No \\
\hline
\cite{goyal2013distributed} & All nodes & BFD, SBTM, DBTM & Contention-based &  {Decentralized} & Busy tone & Yes (By interference measurement by $\mathtt{SRx}$) \\
\hline
Janus\cite{kim2013janus} & AP only \newline (UEs are HD) & DBTM & Scheduling-based &  {Centralized} & Centralized scheduling & Yes (By conflict map) \\
\hline
PoCMAC\cite{choi2015power} & AP only \newline (UEs are HD) & DBTM & Mixed (Contention-based but AP assigns $\mathtt{Tx}$ power) &  {Hybrid} & RTS/CTS & Yes (By optimal $\mathtt{Tx}$ power selection) \\
\hline
A-duplex\cite{tang2015duplex} & AP only \newline (UEs are HD) & SBTM & Mixed (Contention only at uplink) &  {Hybrid} & RTS/CTS, Busy tone & Yes (By SIR map) \\
\hline
\cite{chen2017probabilistic} & AP only \newline (UEs are HD) & SBTM & Mixed (Contention-based but AP assigns access probability) &  {Hybrid} & SNR broadcast & Yes (By uplink power control) \\
\hline
MaSTAR\cite{kim2017asymmetric} & AP only \newline (UEs are HD) & DBTM & Mixed (Contention only at uplink) &  {Hybrid} & RTS/CTS, Busy tone & Yes (By link map) \\
\hline
FECS-MAC\cite{liu2019hidden} & All nodes & BFD, DBTM, SBTM & Contention-based &  {Decentralized} & Secondary carrier sensing & Yes (By power exchange info) \\
\hline
 {\cite{marlali2016s} }& { AP and a subset of nodes }& { BFD }& { Contention-based }&  {Decentralized} & { No }& { No }\\
\hline
 {\cite{aijaz2017simultaneous} }& { AP and a subset of nodes }& { BFD, SBTM }& { Mixed (Contention-based but AP selects secondary user) }&  {Hybrid} & { RTS/CTS }& { Yes (By neighbor map) }\\
\hline
 {MASTER\cite{aijaz2019exploiting} }& { AP and a subset of nodes }& { BFD, SBTM }& { Mixed (Contention-based but AP selects secondary user) }&  {Hybrid} & { Centralized secondary user selection }& { Yes (By RSSI table) }\\
\hline
\end{tabularx}
}
\end{table*}

\begin{figure*}[t]
  \centering
  \includegraphics[width=1.8\columnwidth]{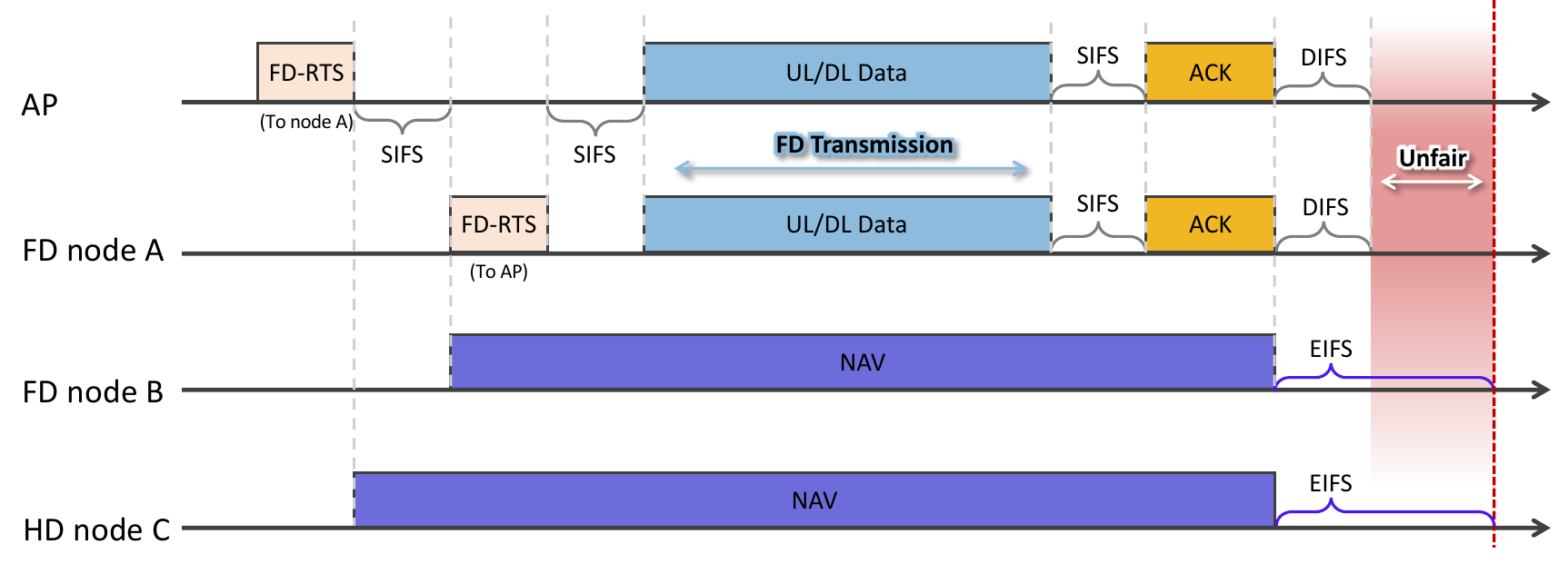}
  \caption{ {An illustration depicting the fairness issue between FD and HD nodes arising from the EIFS waiting time. In this scenario, an AP and FD node A are engaged in FD transmission, with FD node B located near node A and HD node C in the vicinity of the AP.}}
  \label{fig:fd_fairness}
\end{figure*}

\subsubsection{Trade-off between full-duplex link opportunity and fairness}

The trade-off between FD link opportunity and fairness is a major challenge in FD MAC. In an FD system, some nodes are more favorable as a $\mathtt{SRx}$ due to having fewer neighbors and less INI. However, always selecting these nodes as the $\mathtt{SRx}$ may lead to unfairness among the FD nodes. This is similar to the fairness issue in HD communications, where always selecting nodes with good channels as receivers may increase overall transmission efficiency but cause starvation. 

The issue of unfairness can persist even when FD and HD nodes are given equal contention probabilities. {Fig.~\ref{fig:fd_fairness} illustrates the example scenario of the fairness with the coexistence of FD and HD nodes.} This is because HD nodes cannot detect whether FD links have been established among their neighboring nodes, and instead interpret ongoing FD transmissions as collisions. As a result, they must wait for an extended inter frame space (EIFS) time, which is longer than the typical distributed inter frame space (DIFS) wait time in errorless situations, or double their contention window, leading to an increase in their average wait time~\cite{duarte2013design}. This can cause unfairness, as FD nodes have a higher chance of accessing the medium during the extended wait times of HD nodes, leading to reduced overall network performance.

While a system that prioritizes FD communication over HD communication can improve spectral efficiency, it may come at the cost of reduced fairness. Conversely, giving equal transmission opportunities to both FD and HD nodes may not fully utilize the potential advantages of FD communication. Therefore, the FD MAC algorithm should consider the number of FD and HD nodes present in the network and use a scheduling strategy that balances the trade-off between FD link opportunity and fairness. 

\subsection{Existing Full-Duplex MAC Protocols}

In this section, we provide a brief classification of selected FD MAC protocol designs. As in Table~\ref{tab:classification_fdmac}, we classify the exisiting work on FD MAC based on following criteria:
\begin{itemize}
\item FD node types and channel access methods
\item Hidden nodes mitigation methodology
\item Alleviation method of inter-node interference
\item  {Methodology for accounting for legacy HD nodes}
\item  {Approach for guaranteeing fairness}
\end{itemize}
\subsubsection{FD node types and channel access methods}
One way to classify FD MAC designs is based on their assumptions about the availability of FD capability for each node and how channel access is established. Some works in FD MAC, such as {~\cite{sahai2011pushing, singh2011efficient, duarte2013design, goyal2013distributed, liu2019hidden, marlali2016s}}, focused on BFD transmission or ad-hoc networks and treat all nodes as FD nodes. Due to their distributed properties, they tended to rely on contention-based channel access methods that modify conventional CSMA/CA protocols. These methods typically required additional bits for FD to the 802.11 MAC header and introduced various techniques like probabilistic buffer reordering or batch acknowledgement  (ACK) to support efficient FD transmission. In those contention-based methods, fairness between FD and HD nodes was considered by making FD nodes wait for some time slot~\cite{sahai2011pushing, duarte2013design} or selecting secondary links considering fairness~\cite{singh2011efficient, kim2020performance}.

One of the biggest challenges in those ad-hoc networks is asynchronous FD link establishment~\cite{singh2011efficient}. When there is ongoing transmission between two nodes, a node may want to establish another transmission link to one of the nodes, fully utilizing their FD capability. The problem is that when the $\mathtt{PRx}$ is already receiving packets, it cannot estimate its SI channel as it is already used to receive packets from the $\mathtt{PRx}$. To alleviate this issue, one may restrict the initiation of FD links to the $\mathtt{PRx}$~\cite{sahai2011pushing}, or only permit the establishment of secondary links to the preparation stage of the primary transmission (i.e., control signal exchange stage) to ensure that every transmission is synced and required information (e.g., estimated SI channel) can be prepared before data transmission~\cite{goyal2013distributed, kim2020performance}. However, these methods inhibit the full utilization of the FD capability and should be studied continuously in the future.

Recently, there has been an increasing trend in focusing on an AP-STA structure~ {\cite{kim2013janus, choi2015power, tang2015duplex, chen2017probabilistic, kim2017asymmetric, kim2020performance, aijaz2017simultaneous, aijaz2019exploiting}},  {where the APs have full FD capability, and the STAs can either operate in HD or FD mode.} These assumptions are reasonable since FD capability requires complex SIC or multiple transmitters or receivers, which increase complexity and power consumption and are lacking in typical mobile devices. One advantage of these assumptions is that there is a central node (i.e., AP), allowing for centralized channel access methods.

For example, Janus~\cite{kim2013janus} had a centralized, scheduling-based channel access method by the AP, which designated the transmission order and data rate of each node to improve transmission efficiency and enable more complex algorithms for fairness consideration~\cite{kim2013janus, tang2015duplex}. However, this centralized scheduling required more control signal overhead (e.g., regular scheduling order broadcasting), which can reduce FD transmission opportunities. To balance control overhead and efficiency, some works had adopted mixed channel access methodologies {~\cite{kim2013janus, choi2015power, tang2015duplex, chen2017probabilistic, kim2017asymmetric, aijaz2017simultaneous, aijaz2019exploiting}.} These methods were essentially contention-based, but they incorporated different access probabilities~\cite{chen2017probabilistic} or transmit power levels~\cite{choi2015power}, or they only used contention for primary transmissions and allowed the AP select the $\mathtt{SRx}$~ {\cite{tang2015duplex, kim2017asymmetric, aijaz2017simultaneous, aijaz2019exploiting}.}


\subsubsection{Hidden nodes mitigation methodology}
To mitigate hidden nodes, most existing FD MAC designs used RTS/CTS broadcasting, similar to HD MACs {~\cite{sahai2011pushing, duarte2013design, choi2015power, tang2015duplex, kim2017asymmetric, kim2020performance, aijaz2017simultaneous},} or transmitted a busy-tone utilizing their ability to transmit and receive simultaneously~\cite{singh2011efficient, goyal2013distributed, tang2015duplex, liu2019hidden}. Some approaches aimed to utilize these collision-alleviating broadcasts to convey useful information such as SNR~\cite{chen2017probabilistic} or eliminated the need for hidden node mitigation by using centralized scheduling~\cite{kim2013janus}. FECS-MAC~\cite{liu2019hidden} eliminated the need for hidden node mitigation by utilizing a secondary carrier sensing method. Here, the $\mathtt{STx}$ senses first, using its FD capability, to compute whether a collision will occur based on its transmission power and the neighboring nodes' interference level, based on the proposed ellipse interference model.  {MASTER~\cite{aijaz2019exploiting} avoids the requirement for hidden node mitigation by letting the AP to select the secondary user through CSMA/ECA's deterministic backoff mechanism in conjunction with the RSSI table.}

\subsubsection{Alleviation method of inter-node interference}

There were several INI alleviation methods used in existing FD MAC designs. Early FD MAC works addressed the INI problem by blocking secondary transmission that might cause interference between neighboring nodes. For example, FD-MAC~\cite{sahai2011pushing} avoided INI by allowing each node to estimate the network topology by snooping MAC header of other nodes and refraining from transmitting if the node that makes clique with itself is currently transmitting/receiving. ContraFlow~\cite{singh2011efficient} restricted initiation of secondary transmission to only the $\mathtt{PRx}$. While these restrictions could effectively resolve the INI problem, they might have missed possible FD opportunities due to overly restrictive transmission rules.

Recent FD MAC designs {~\cite{goyal2013distributed, tang2015duplex, kim2017asymmetric, liu2019hidden, kim2020performance, aijaz2017simultaneous, aijaz2019exploiting}} addressed INI by using the signal-to-interference ratio (SIR) concept. {~\cite{goyal2013distributed, aijaz2017simultaneous, aijaz2019exploiting}} introduced SIR-based decision making for establishing secondary links, while subsequent studies developed the idea further by maintaining an SIR map for efficient scheduling~\cite{kim2013janus} or using it to select secondary receivers based on transmission efficiency~\cite{tang2015duplex, kim2017asymmetric, liu2019hidden}. SIR information was collected through periodic reply signals from STA nodes {~\cite{kim2013janus, aijaz2017simultaneous, aijaz2019exploiting}}, power exchange packets~\cite{liu2019hidden}, or bits in RTS packets~\cite{tang2015duplex}. Additionally, some designs managed a link success rate table, which implicitly considered fairness and INI simultaneously~\cite{kim2020performance}.  {\cite{aijaz2019exploiting} proposed to dynamically adapt carrier sensing threshold to maximize the FD communications opportunity.}

There were also approaches in existing FD MAC designs that optimized transmit power to balance the trade-off between higher link speed and lower INI. For instance, in~\cite{choi2015power}, received signal strength-based contention window adjustment was conducted to favor nodes with higher signal strength and lower interference from neighboring nodes. Additionally, the AP solved a power optimization problem to maximize the sum spectral efficiency given minimum SINR constraints. In~\cite{chen2017probabilistic}, the $\mathtt{PTx}$ adjusted its UL power to minimize its effect on neighboring $\mathtt{SRx}$, by ensuring that the SINR of the AP was higher than a threshold, which reduced the burden on the AP to calculate the SINR value of every pair to determine an appropriate modulation scheme. The necessary information to optimize transmit power was collected by control frames such as RTS/CTS headers~\cite{choi2015power, chen2017probabilistic}. These methods allowed for better control over the transmission power, and could be used to achieve more efficient and reliable transmission while minimizing interference to other nodes.


\subsubsection{ {Methodology for accounting for legacy HD nodes}}

 {One essential aspect of FD MAC is backward compatibility. As network systems evolve slowly, and FD capability is both costly and computationally heavy, FD MAC protocols must consider the coexistence of FD and legacy HD nodes. A typical approach to ensuring backward compatibility is to build upon the widely-adopted 802.11 MAC standard with minimal modifications. For instance, \cite{marlali2016s, aijaz2017simultaneous, aijaz2019exploiting} leverage reserved bits in the 802.11 header to identify FD capability or broadcast the transmission type (FD or HD). Since legacy nodes are agnostic to the contents of these reserved bits, this approach guarantees backward compatibility.}

 {\cite{aijaz2017simultaneous, aijaz2019exploiting} also address the issue of ACK timeouts when FD and HD nodes coexist. Consider a scenario where node A is transmitting to the AP while the AP simultaneously sends data to node B. If node A finishes its transmission earlier than node B, the AP may struggle to transmit the ACK message at the right time because it's still sending data to node B. This leads to ACK timeouts, triggering retransmissions even when there are no transmission errors. To mitigate this problem, the proposed solution involves delaying the transmission start time or appropriately setting the ACK timeout timer.}

\subsubsection{ {Approach for guaranteeing fairness}}
\label{sec:fairness}

 {
Fairness is another critical concern in MAC protocols. Like conventional MAC protocols, FD MAC protocols account for fairness using metrics such as the number of successive transmissions~\cite{singh2011efficient} for link selection or employ techniques like deficit round robin~\cite{kim2013janus, tang2015duplex}. However, due to the coexistence of FD and HD nodes, new fairness issues arise, especially concerning the EIFS waiting time that legacy nodes endure when sensing collisions or ongoing FD transmissions. Ensuring fairness between HD and FD nodes, possibly without any modification to legacy nodes, is crucial. Various solutions exist, such as centralization-based schemes~\cite{singh2011efficient, kim2013janus, tang2015duplex, kim2020performance}, which can consider these fairness issues and provide more opportunities for legacy HD nodes. Enforcing FD nodes to wait for EIFS time rather than DIFS~\cite{duarte2013design} can be a solution, albeit at the expense of efficiency. Signaling-based approaches~\cite{aijaz2017simultaneous} can also address fairness issues while retaining efficiency. However, these solutions often require modifications to legacy HD nodes, potentially introducing fairness issues for non-updated legacy nodes.}


  \begin{table*}[htbp]
{\footnotesize
\caption{Advantages, Drawbacks, and Key Features of Existing Full-Duplex Medium Access Control Designs}
\label{tab:features_fdmac}
\centering
\def\arraystretch{1.5}
\begin{tabularx}{\textwidth}{>{\hsize=0.3\hsize}X|>{\hsize=1\hsize}X|>{\hsize=1.1\hsize}X|>{\hsize=1.6\hsize}X}
\hline
\textbf{Reference} & \textbf{Advantages} & \textbf{Drawbacks} & \textbf{Key Features} \\
\hline
FD-MAC\cite{sahai2011pushing} & Enhances FD opportunities through probabilistic non-head-of-line packet transmission, and prevents channel domination by FD nodes & Necessitates an additional 45-bit FD header, and misaligned timers may cause synchronization issues in the shared random backoff & Introduction of additional FD header bits, shared random backoff to mitigate starvation, header snooping for network topology estimation, and probabilistic buffer reordering to enhance FD opportunities \\
\hline
ContraFlow\cite{singh2011efficient} & Addresses fairness among nodes, and employs asynchronous transmission to enhance efficiency & Conditions to initiate secondary transmission are limited; SI channel estimation issue during reception are not addressed & Exclusive initiation of secondary transmission by $\mathtt{PRx}$ to enable async transmission; Weighted (link success probability-based) $\mathtt{SRx}$ selection to avoid secondary collisions; Pressure based link selection that prioritizes least-used and shorter links for fairness \\
\hline
\cite{duarte2013design} & Modifies the existing 802.11 standard minimally for seamless adaptation, and addresses fairness between FD and HD nodes & Focuses solely on BFD, has higher complexity due to distinct strategies for FD-only and mixed FD and HD scenarios & Prompt data transmission of $\mathtt{Rx}$ after CTS with buffer inspection; transmission of ACK prior to its reception to prevent deadlock; FD transmitting nodes wait for EIFS instead of DIFS to mitigate unfairness between FD and HD nodes \\
\hline
\cite{goyal2013distributed} & Addresses all types of full-duplex (BFD, SBTM, DBTM), features minimal additional FD header bits (3 bits), and considers SIR for node selection & Employs a busy tone approach, and assumes accurate bit decoding during busy tone of another nodes & Header bits employed to indicate the type of FD link or broadcast reception willingness, which mitigates hidden node issues; Packet snooping implemented for SIR measurement; Random subcarrier-based contention adopted in SBTM \\
\hline
Janus\cite{kim2013janus} & Eliminates busy tone/random backoff using centralized approach for improved transmission efficiency, efficiently measures interference with a single control signal, and takes asymmetric link into account & Ordered probe request response could introduce additional overhead, FD transmission opportunities might be missed due to slot registration contention, and batch ACK may lead to increased retransmission latency & SIR-based conflict map managed by AP to determine FD data rate; ordered signaling with a unique ID for each node, batch ACK employed to avoid interference, deficit round robin-based fairness consideration based on channel access time; queue rate-based heuristic utilized to identify scheduling strategy with the shortest transmission time \\
\hline
PoCMAC\cite{choi2015power} & Enhances spectral efficiency through AP-assigned optimal $\mathtt{Tx}$ power strength & Introduces increased overhead due to CTS-U frame (utilizes 6 bytes per contention candidates), and places a burden on clients to overhear control signals & Received signal strength-based $\mathtt{Rx}$ contention; Optimal $\mathtt{Tx}$ power adjustment to maximize sum rate considering INI; INI map calculation and $\mathtt{Tx}$ power notification by AP \\
\hline
A-duplex\cite{tang2015duplex} & No need of busy tone signal for clients; Fair header overhead (additional 8 bits for RTS frame), Considers asymmetric link & Outdated SIR map can degrade performance & Dual-link detection using CTS duration value; SIR map calculation based on Received Signal Strength Indicator (RSSI) to determine $\mathtt{SRx}$ (by AP); Fairness and FD opportunity both considered by virtual deficit round robin \\
\hline
\cite{chen2017probabilistic} & Exhibits reduced collision probability due to probability-based contention window and places minimal scheduling burden on the AP & Necessitates STAs to broadcast their SNR before DL reception, and demands transmission of channel access probability to each STA & Epoch-based scheduling considering average traffic demands; AP-assigned channel access probability for each node, which takes into account fair share and total throughput maximization; Probability-based backoff; UL power control to mitigate INI \\
\hline
MaSTAR\cite{kim2017asymmetric} & Measures SINR efficiently utilizing FD capabilities of the AP and streamlines block ACK transmission through partial overlap & Requires SINR measurement report from STAs, leading to additional overhead & $\mathtt{PTx}$ identification of AP by header snooping or RTS/CTS packet; AP-managed Link SINR map to coordinate secondary transmission; Partially overlapped block ACK \\
\hline
FECS-MAC\cite{liu2019hidden} & Addresses all types of FD link (BFD, SBTM, DBTM) and employs simultaneous carrier sensing using FD capabilities of each node to enhance spatial reuse & Requires periodic exchange of power set information & Concurrent carrier sensing by $\mathtt{STx}$ to avoid INI; carrier sensing power threshold determined using an ellipse model to mitigate INI; reception power measured by each node through power exchange packets \\
\hline
\cite{kim2020performance} & Facilitates efficient SIR data calculation through RTS/CTS overhearing; incorporates fairness algorithmically in $\mathtt{SRx}$ selection & Requires additional overhead for SINR measurement reports from STAs & FD pair table managed by AP to select $\mathtt{SRx}$; $\mathtt{SRx}$ selection considering fairness (i.e., accounting for the number of packets received and the probability of a given node being chosen for $\mathtt{SRx}$) \\
\hline
 {\cite{marlali2016s}} &  {Modifies 802.11 standard minimally for coexistence with legacy nodes; Introduces a realistic SI model for analysis} &  {Necessitates updates to legacy nodes for FD-HD fairness due to EIFS issues; Vulnerable to synchronization issues arising from hidden nodes} &  {DIFS waiting time for collision detection due to FD transmission; Shared backoff window size-based contention mechanism; Reserved bits for backward-compatible FD capability exchange} \\
\hline
 {\cite{aijaz2017simultaneous}} &  {Achieves backward compatibility through meticulous ACK timeout and contention fairness solutions} &  {Relies on coarse-grained neighbor data compared to SIR map-based strategies; Updates needed in legacy nodes to address EIFS issues} &  {FD node discovery through reserved bits; Neighbor information exchange for INI mitigation; Transmission timing adaptation to resolve ACK timeout issues; Contention fairness by CTS-FD (ignoring collision for FD) or FDTI (sending clear message to prevent EIFS waiting time)} \\
\hline
 {MASTER\cite{aijaz2019exploiting}} &  {Backward compatibility; Eliminates RTS/CTS handshaking; Considers FD  preparation time; Enhances FD opportunities compared to~\cite{aijaz2017simultaneous}} &  {Introduces extra network load due to Neighbor RSSI packets; Offers limited opportunities compared to SIR-based approaches} &  {Removal of handshaking via CSMA/ECD deterministic backoff; Dynamic carrier sensing threshold adaptation for FD opportunity maximization} \\
\hline
\end{tabularx}
}
\end{table*}

For more detailed explanations for each FD MAC method, including advantages, drawbacks, and key features, please refer to Table~\ref{tab:features_fdmac}.

\section{ {User Scheduling Algorithm for \\Full-duplex Cellular Network}}\label{sec.US}
\begin{figure}

	\begin{center}
		{\includegraphics[width=1\columnwidth,keepaspectratio]
			{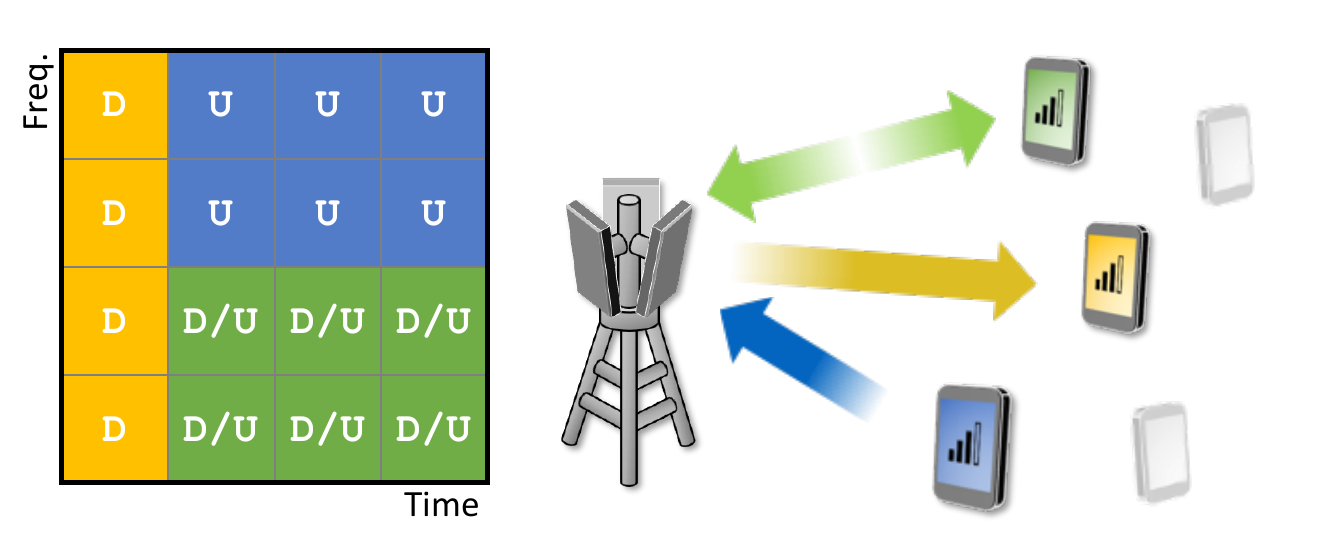}%
			\caption{Scheduling issue for FD network design.}
			\label{fig.scheduling}
		}
	\end{center}

\end{figure}

{The scheduling algorithm in the FD network has been thoroughly researched in the literature for three main reasons. Firstly, the network can achieve FD throughput gain under specific conditions~\cite{FeasibleFDchae, compactFDd2d}. Secondly, the CLI and ICI, such as intra-cell and inter-cell interference levels, increase compared to HD scheduling scenarios. In an FD network, lastly, it is typically assumed that the network includes an FD-compatible BS node and HD-only UE nodes. This user pairing approach results in a scheduling algorithm with a larger search space compared to the HD scheduling algorithm.}

In FD network scheduling, there are four main steps: 1) UL and DL transmission direction assignment or BS node selects the duplex mode. 2) UL and DL band or RE allocation. 3) User selection or user pairing. 4) Power control. An FD network naturally involves coordinating multiple BSs, introducing the concept of network-assisted FD (NAFD). While most studies focus on centralized control, some suggest distributed control.

Three network-level metrics can be used for scheduling: network sum-rate, fairness, and other metrics such as QoS and coverage. The network sum-rate focuses on the advantage of doubling spectral efficiency, considering not only the sum-rate but also the weighted sum-rate of UL and DL. These metrics can be applied to users within a cell or across multiple cells, and various studies~\cite{jointupHD, jointsubpowHD, US-NAFD} applied matching algorithms and heuristic iteration-based methods.

Fairness is another critical metric, incorporating max-min and proportional fairness rates. The max-min fairness rate aims to maximize the minimum rate among users in network scheduling. Fairness control balances the link-level throughput for each UL-UE node, $R^\text{UL}$,  and DL-UE node, $R^\text{DL}$, respectively. The proportional fairness rate, usually used for distributed control, selects the user with the maximum (or minimum value of) the ratio of the window-based average rate, and the predicted rate if a channel is assigned to a UE node such as follows:
\begin{equation}
i[n] = \text{argmax} \frac{R^\text{DL}_i[n]}{\overline{R^\text{DL}_i}[n-1]},
\end{equation}
where $\overline{R^\text{DL}_i}[k]$ is the expectation of $R^\text{DL}$ over time slot index $k$, and the above equation selects DL-UE node $i$ for HD scheduling problem.


\subsection{Conventional Half-duplex User Scheduling}

In the conventional duplex scenario, HD, user scheduling, and power control are the classic optimization problem regarding various network system components. Traditional user selection varied between UL only and DL only scenarios in multi-cell or multi-BS scenarios, but FD considered both aspects~\cite{CoScheOFDMA, 07kianiMax}. 
For DL user selection, the interferers only include BS nodes, making it easy to calculate the channel gain from adjacent cells before the centralized system determines the overall scheduling. In contrast, when it comes to UL user selection, UE nodes in adjacent cells generate ICI. Therefore, the interference or channel gain from ICI, i.e., UE-to-BS, cannot be calculated before the scheduling is determined.

Several approaches consider the centralized network concepts for FD network design. Coordinated multipoint transmission (CoMP)~\cite{202008jsac, FD_CFMM_ICC, 201604wcl, CBF} and beamforming-based approaches~\cite{CBF} were studied. Cloud radio access (C-RAN) was another approach~\cite{muBF, cranMIMO, cranTDD, US-cranFD}, where baseband processing was shifted from the BS nodes or APs to "the cloud." C-RAN of IBFD was also explored~\cite{cranIK}. Complexity analysis, however, remained an essential aspect of C-RAN. Studies also investigated DL network power consumption minimization~\cite{cranMIMO} and the joint design of user selection algorithms and transceivers for TDD~\cite{muBF, cranTDD}. The detailed network concept for FD will be discussed in Section~\ref{sec.CLI}.

Some studies have explored the use of MIMO beamforming with BS nodes, leading to issues like SI, CLI, and ICI degradation~\cite{US-NAFD}. Multi-antenna systems used for beamforming can also result in deafness, which can be mitigated with the help of an FD MAC protocol~\cite{FDmac_Mag}. In the context of interference coordination in HD scheduling, numerous studies have proposed optimization methods and algorithms, based on static frequency resource allocation and dynamic resource allocation~\cite{201310cst}. In FD systems, CLI roughly doubles, necessitating the joint allocation of DL and UL resources to support multiple simultaneous transmissions and receptions.

 {Besides the theoretical work supporting HD networks through advanced scheduling strategies, recent advances in B5G/6G networks also pave the way for adopting FD networks~\cite{3gpp38912}. To support various latency constraints depending on the application, Type A and Type B scheduling are defined as distinct mapping strategies for physical channels. While Type A scheduling involves less frequent scheduling decisions, Type B scheduling is designed to provide dynamic resource allocation for users through the flexible adjustment of scheduling parameters~\cite{sch202112csm}. This flexibility extends to more varied options in determining the start symbol and allocation length in symbols, achieved by adaptively dividing slots into mini-slots. Given that inefficient user scheduling is often a major bottleneck for agile services, these enhancements in the flexibility of user scheduling are crucial for the practical implementation of FD network optimization~\cite{sch202112csm,sch202306jsac}.}

\subsection{General Problem Formulation of User Scheduling}

In an IBFD scenario, as most research has considered this as the system model, user scheduling is simplified to focus primarily on user pairing and power allocation. User paring is defined as $\{\pi_b(i)\}$ for each BS node $b$ for DL-UE node $i\in \mathbb{U}_b$. User pairing indicator $\pi_b$ designates the index of a UL-UE node as $j=\pi_b(i)$ for a DL-UE node $i$. This type of pairing problem results in the necessity of a combinatorial optimization problem, the complexity of which significantly exceeds that of a polynomial, with a complexity of $\mathcal{O}(|\mathbb{U}_b|!)$ for a specific cell $b$.

Controlling the values of ${p^\text{UL},p^\text{DL}}$ relates to the power allocation problem. In the case of HD, it is advisable to increase the transmit power to optimize the SINR of the desired signal, and apply criteria like maximum ratio transmission (MRT). However, in an FD network, as the SI and CLI increase and affect the SINR for both DL and UL, the need for an appropriate power allocation criteria becomes evident. If a MIMO antenna is configured, the channel gain $G$ can also vary depending on the node's location. By integrating user pairing and power control, a thoughtful user selection approach reduces the need to forcefully reduce power to mitigate SI and INI. A general user scheduling problem is defined as follows:
\begin{equation}
\begin{aligned}
		\underset{\{\pi_b\}, \{p^\text{DL}\}, \{p^\text{UL}\}}{\text{maximize}} \quad &\underset{b\in\mathbb{B}}{\sum}\underset{i\in\mathbb{U}_b}{\sum}\left\{w\left(R^\text{DL}_i\right)+w\left(R^\text{UL}_j\right)\right\}\\
		\textrm{subject to} 
		\quad\quad &0 \leq p^\text{DL}_{b} \leq P^\text{DL}_\text{max},\\
		\quad\quad &0 \leq p^\text{UL}_{j} \leq P^\text{UL}_\text{max},\\
		\quad\quad & {R_i^\text{DL},R_j^\text{UL}\geq R_\text{threshold}},\\
		\quad\quad & \forall i,j\in\mathbb{U}_b, \forall b\in\mathbb{B},
\end{aligned}
\end{equation}
 {
UL-UE node typically operates with a lower transmit power compared to DL-UE node, denoted as $P^\text{UL}_\text{max} \leq P^\text{DL}_\text{max}$. This leads to challenges in canceling CLI, often resulting in a lower SINR for the UL, symbolized as $\text{SINR}^\text{UL}_j$.  In any UL link scenario, a UL-UE node, denoted as $j\in \mathbb{U}_b$, sends a signal to its designated BS node $b$. This BS node $b$ encounters SI, represented as $ I^\text{SI}_\text{BS} = p_b^\text{SI} G_b^\text{SI} $, and also faces ICI from another BS node $b'$, which is described as $I_{\text{BS}\leftarrow \text{UE}} = \sum p_{b'}^\text{DL} G_{b'}^\text{BU}$. In comparison to the level of interference it faces, the signal of interest for the BS node, specifically $S^\text{UL}_{\text{BS}\leftarrow \text{UE}} = p_{j}^\text{UL} G_{b,j}^\text{BU}$, is relatively weak. Consequently, this results in a less favorable $\text{SINR}^\text{UL}_j$ and may lead to outages during the network scheduling process. Especially, SINR degradation of UL link leads fairness and coverage loss for the BS node~\cite{XDD}.}

 {In a DL, despite the BS node having sufficient transmit power, it faces significant level of interference as ICI and CLI. When a BS node $b$ services with a DL-UE node $i$, ICI arises from an adjacent cell's BS node $b'$, denoted as $I_{\text{BS}\leftarrow\text{BS}} $. CLI, in particular, is a critical concern. It occurs when an UL-UE node $j $ interferes with the DL-UE node $i$, generating interference $ I^\text{CLI}_{\text{UE}\leftarrow \text{UE}} = p_{j}^\text{UL} G_{i,j}^\text{UU}$. A major problem arises if the DL-UE and UL-UE are located too closely, leading to an increase in $G_{i,j}$ and a substantial rise in CLI. This results in a decrease in the SINR for the DL, specifically $\text{SINR}^\text{DL}_{b\rightarrow i}$, becoming unfavorably low. To mitigate this issue, an effective user pairing algorithm is essential for proper CLI management between DL-UE and UL-UE nodes to maintain sufficient SINR levels.} 
 
{Due to these issues, it is natural that while distributing power to each node, the QoS of certain nodes can degrade significantly. In some cases, it is beneficial for the network to avoid allocating power to links with unfavorable channel conditions, including desired link and interference. Hence, it is vital to incorporate conditions into the network optimization process to guarantee that each node fulfills the QoS requirements. Ensuring a minimum UL/DL rate, denoted as $R_\text{threshold}$, is necessary as a criterion throughout the scheduling process.
}

 {The determination of the function $w(R)$ in terms of UL/DL throughput, $R$, means the obejctive of the network-level performance. Types of metrics include sum-rate, weighted sum-rate, proportional sum-rate, and max-min sum-rate~\cite{Fair_CST,Fair_JSAC,Fair_TIT}.} 
\subsubsection{ {Throughput maximization}}  {When $w(R)=R$, the optimization problem becomes the network sum-rate maximization. The sum-rate maximization strategy in primarily aims to maximize total network throughput, often disregarding the equitable distribution of resources among nodes. Its main goal is to achieve the highest overall data rate, which can result in users with better channel conditions or closer proximity to a BS node receiving a larger share of resources. }

 {From an algorithmic perspective, sum-rate maximization typically involves solving complex linear or nonlinear optimization problems that aim to maximize the aggregate user throughputs. The primary challenge lies in the complexity of these optimization problems, particularly in large-scale networks.}
\subsubsection{ {Fairness maximization}}  {In cellular networks, two qualitative fairness approaches are commonly used: proportional fairness and max-min fairness. Proportional fairness is achieved by maximizing the sum of log rate, symbolized as $w(R)=\log(R)$. Proportional fairness strikes a balance between overall network throughput and fairness. It aims to ensure that any gain in resource allocation for one node is not more significant than the loss for another, often resulting in higher network throughput than max-min fairness.}

 {On the other hand, max-min fairness focuses on the equitable distribution of resources, as $w(R)=\text{min}\{R_1,\cdots,R_i\}$. Its primary objective is to maximize the minimum rate received by any node, ensuring the fairest possible allocation. This approach can be challenging as it involves determining the minimum rate and then reallocating resources in a way that balances rates across nodes without reducing others' throughput.}

\subsection{Computational Approaches: Scheduling Algorithm and Complexity}

Various optimization challenges and approaches arise in the context of FD network user scheduling problems. Generally, the optimization problem is non-convex, nonlinear, and mixed discrete, making it difficult to find a global optimal scheduler~\cite{US-PowMulti, US-alpha}. In some cases, the optimization problem may lead to NP-hard scenarios~\cite{US-NAFD, US-MMF}. This complexity also extends to HD scheduling problems.

To tackle this issue, researchers often decomposed the problem into multiple stages~\cite{US-NAFD, US-MMF, US-alpha}, including user scheduling and power or resource allocation. Various strategies have been suggested to confront these challenges in FD network user scheduling problems. Stochastic geometry-based approaches~\cite{alpha1, 15leeHybrid} were used for FD multi-cell systems, although they have yet to consider multi-UE power control. Graph theory-based methods were also applied to address scheduling issues, with matching algorithms deployed. One effective technique for addressing non-convex scheduling problems is geometric programming (GP)~\cite{boyd2007tutorialGP, 07chiangPowGP}. GP is particularly beneficial for considering fairness metrics, such as the sum of log rates, and offers an efficient optimization of objective functions with polynomial or monomial expressions.
 Another strategy employs to tackle non-convex optimization issues was successive convex approximation (SCA)~\cite{SCA, 13razUniSCA}. SCA simplified these problems by converting them into a series of convex subproblems, providing scalability and convergence properties. Although SCA hard to guarantee a global optimum, it often converges to a stationary point or locally optimal solution. This solution is typically satisfactory for practical wireless communication problems, especially when a good initial point is provided or the non-convexity is relatively mild. In these optimization approaches, it was assumed that the BS node had knowledge of channel gains for all UE nodes through channel state information (CSI) reporting, even the BS node was aware of the channel between all UE pairs.

\subsection{Effect of Interference Cancellation on Full-duplex User Scheduling}
The impacts of imperfect SIC on user scheduling and power control at the network level are observed across various scenarios. The performance disparity among networks is influenced by the degree of achievable SIC. In general, SI is suppressed by SI channel gain (analog SIC), and its residual impact is reduced to noise levels through digital SIC~\cite{US-alpha, IanBSI}. Concerning sum-rate or spectral efficiency, both UL and DL rates deteriorate as the performance of SIC fluctuates. The degradation of UL and DL rates is more pronounced when a significant disparity exists between the BS node and UE node transmit power. The disparity of transmit power is defined as follows:
\begin{equation}
\delta=P^\text{UL}_\text{max}/P^\text{DL}_\text{max}.
\end{equation}
In cases of lower transmit power disparity, the scheduler selected FD mode, leading to an increase in CLI and ICI~\cite{US-alpha}. In environments where analog beamforming is feasible, the SI channel gain varied depending on the direction of the UE pair~\cite{IanBSI}. Some studies measured spectral efficiency and the number of users a network could accommodate concerning the cancellation of ICI, mainly focusing on suppression ability of BS-to-BS ICI ranging from $10$ to $15$~dB~\cite{US-NAFD}.

Regarding fairness, the convergence to the target scheduler was slower for proportional fairness in scenarios with SIC at $95$~dB compared to perfect SIC scenarios~\cite{US-PowMulti}. 
The weaker the SIC, the fewer BSs that chose FD mode~\cite{US-PowMulti}. While SIC ranged from $110$~dB to $60$~dB, the max-min fairness rates of various methods declined~\cite{US-MMF}. The overall SI in the network is reduced due to SIC technology, providing the network with additional capacity to manage interference. As a result, the network scheduled UE nodes and power allocation to address the slightly increased interference.

\subsection{ {Full-duplex Heterogeneous Network and User Scheduling}}
 {The integration of FD networks with HetNets markedly enhances their benefits. HetNet is characterized by a combination of a macro BS node and a smaller BS node. This setup often extends to BS nodes with multi-tier transmit power serving UE nodes. The smaller BS-nodes, which transmit relatively lower power, contribute to reduced SI and ICI, thereby facilitating easier network interference management. Notably, the performance of these small BS-nodes in FD mode is enhanced with greater spatial density~\cite{HN_Jemin,HN_Yu}.}

 {Moreover, FD HetNet significantly improves network efficiency by utilizing the same spectrum for both UL and DL, thereby reducing the size of interference and simplifying spectrum management. This combination of FD and HetNet not only doubles spectral efficiency but also leads to overall throughput gain and energy efficiency gains from a network perspective~\cite{HN_Zhang}.}

 {In terms of scheduling in FD HetNet, the process involves UL/DL association for each small BS-node. The two main methods employed are downlink/uplink coupled (DUCo) and downlink/uplink decoupled (DUDe), with the DUDe user association method demonstrating better performance in HetNet when assuming FD~\cite{HN_wong}. Scheduling in FD HetNet is based on certain matching algorithms~\cite{HN_Chen, HN_Dai}, and the small BS-nodes are capable of hybrid functionalities, including duplex mode selection, user association, and power control~\cite{HN_Zhang, heteroFD}. Furthermore, the approach of determining user pairing for multiple small BS-nodes in HetNet can be extended to multi-cell scheduling strategies.}

\subsection{ {Machine Learning Techniques for Full-duplex User Scheduling}}

 {In FD scheduling, channel estimation is crucial, involving channel gain and interference strength measurements. Despite the complexity, machine learning (ML) techniques are being explored for low-complexity, distributed solutions to this NP-hard scheduling problem, diverging from traditional communication approaches.}

 {FD scheduling is unique due to the necessity of user pairing, which entails selecting two users from possible $N^2$ combinations for $N$ UE nodes. ML allows for the simultaneous maximization of sum-rate and fairness, an advancement over traditional methods that typically focus on a single metric $w$~\cite{ML_SVM,ML_spatial}. ML methods also develop site-specific models for BS-node areas, learning appropriate feedback mechanisms from UE nodes.}

 {Two primary ML approaches in FD scheduling are reinforcement learning (RL) and neural network (NN)-based methods. The RL approach, mainly distributed Q-learning, enables nodes to select transmission targets in rapidly changing channel conditions~\cite{ML_dist}. Though this requires substantial CSI information, it facilitates quick scheduling decisions without separating BS and UE nodes. Integrating with earlier discussed HetNet perspectives, multi-agent deep reinforcement learning (DRL) is applied in scenarios where unmanned aerial vehicles act as small BS nodes for UL/DL user pairing~\cite{ML_multi}.}

 {The NN approach focuses on learning UE node locations through Deep Neural Networks (DNN) and proposes feedback based solely on geographic location~\cite{ML_spatial}. This method maximizes sum-rate and ensures fairness by providing multiple solution options. Additionally, it learns interference-related information from spatial data using convolution filters. The flexibility of NNs is further enhanced by pointer networks, which process sequential data and accommodate variable UE-node counts without altering hyperparameters~\cite{ML_pointer}.}

 {ML-based FD network scheduling predominantly employs RL in distributed methods and NN in centralized approaches. These methods address CLI among potential INIs and tackle the complex combinatorial optimization inherent in FD networks, where nodes can function as transmitters, receivers, or both. Through ML, various metrics are maximized, addressing the intricacies of FD network scheduling.}

\section{Cross-link Interference Handling issue for implementing Full-duplex Network}\label{sec.CLI}

\subsection{Cross-link Interference in Cellular Network}

\begin{figure}[t]
	\begin{center}
		{\includegraphics[width=0.9\columnwidth,keepaspectratio]
			{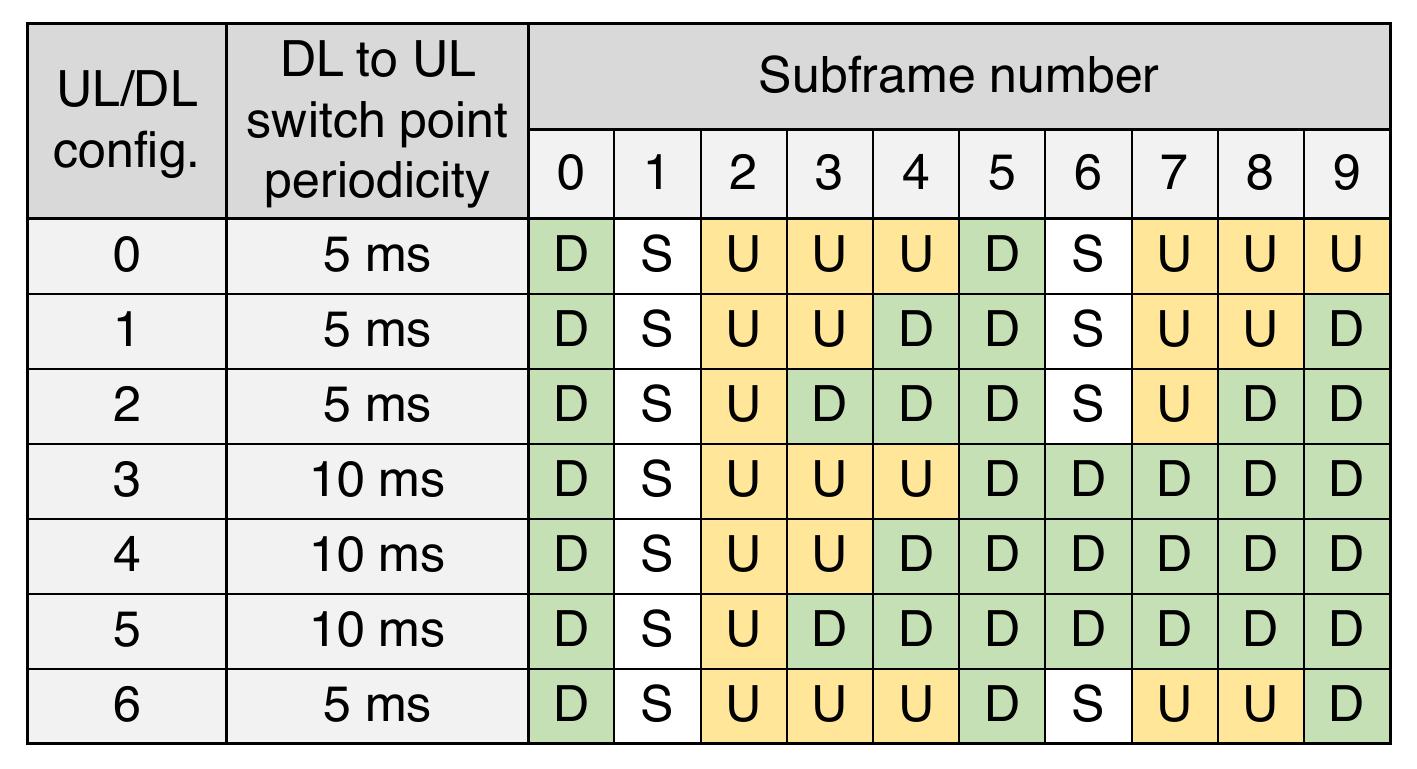}%
			\caption{UL/DL configuration for fixed TDD in LTE.}
			\label{fig01}
		}
	\end{center}
	\vspace{-15pt}
\end{figure}	

\begin{figure*}[t]
	\begin{center}
		{\includegraphics[width=2\columnwidth,keepaspectratio]
			{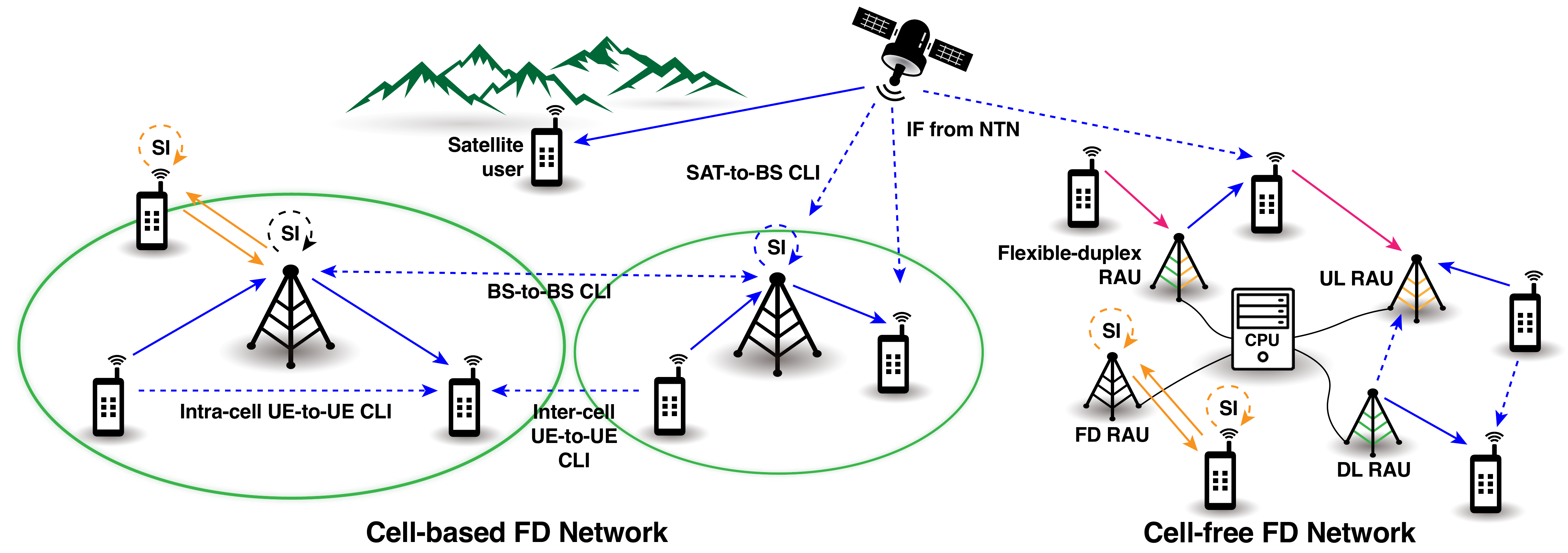}%
			\caption{Comparison of cell-based/-free network structures for FD.}
			\label{fig02}
		}
	\end{center}
	\vspace{-15pt}
\end{figure*}


Flexible and aggressive resource allocation schemes of FD networks offer the potential to utilize limited resources efficiently. As an initial step, 5G NR supported dynamic TDD, enabling distinct UL/DL subframe configurations for different cells~\cite{5gtechnology3gppnewradio}. This flexibility allows adaptation to asymmetric traffic demands by adjusting the switching point of UL/DL slots~\cite{2013icc,2013globecom,2015vtc} and even determining configurations at the symbol level, thus supporting low-latency services. However, the unconstrained allocation of UL/DL configurations introduces CLI, resulting from opposite transmission directions (UL/DL) between neighboring BS nodes~\cite{202010cst,202108wcom}. BS-to-BS CLI, or ICI, occurs when DL-BS node signals interfere with UL-BS nodes, while UE-to-UE CLI arises when the UL-UE node interferes with the DL-UE node. CLI can occur even in conventional TDD systems with identical neighboring BS subframe configurations due to imperfect synchronization and signal delays from adjacent cells~\cite{2016vtc}. {Fig.~\ref{fig01} illustrates the UL/DL configuration in conventional LTE system.} 
Due to the rise of the CLI, coordinated mechanisms for enhanced interference mitigation and traffic adaptation (eIMTA) are thus essential for fully utilizing the flexibility of future communication systems.

A BS-to-BS CLI can significantly impair UL performance, as BSs typically transmit at high power. Macro BSs with high transmission power and numerous antennas are particularly detrimental in this aspect, as the relative power of CLI can be much larger than the desired UL signal. Conversely, UE-to-UE CLI degrades the DL performance of the network. Although UE-to-UE interference is generally smaller than ICI from other BS nodes due to lower transmit power at UE, it can still exceed the power of ICI from other BS nodes depending on the location of the desired and interfering UE nodes~\cite{2010vtc}. As the demand for cellular networks increases to support a growing number of users, deploying denser BS nodes with more antennas is essential. However, this increase in cell density inevitably leads to CLI. As a result, the development of intelligent CLI mitigation techniques is actively being explored in these systems.

\subsection{CLI Handling Strategies in Dynamic TDD Network}
	
Before discussing the CLI handling approach in FD networks, we first introduce CLI handling techniques in dynamic TDD systems.
Since CLI occurs across cells, CLI measurement and joint optimization of resources through network coordination are required. To this end, information is exchanged between gNBs over Xn and F1 interfaces, which can be used for making centralized or distributed decisions for the operation in each cell~\cite{202108wcom}. By leveraging the exchanged information, the network can implement CLI mitigation techniques to effectively manage interference and maintain optimal network performance.

The CLI mitigation mechanisms can be categorized into cell clustering, UL/DL configuration allocation, resource scheduling, and interference cancellation. Cells can be clustered to minimize the BS-to-BS CLI or ICI channel strength from other clusters~\cite{tr36828,2014icc,201809twc2}. In this method, BSs within the same cluster adhere to identical and synchronized UL/DL configurations to avoid CLI. Secondly, although the primary objective of dynamic UL/DL configuration is to accommodate asymmetric UL/DL demands with minimal latency, CLI should also be considered for enhanced system throughput. Researchers have proposed various algorithms that determine UL/DL configurations while suppressing the impact of CLI~\cite{201612jsac,201806wcl}. Furthermore, the coordination of BS nodes enables the implementation of proper user scheduling, resource allocation, power control, and beamforming to mitigate CLI~\cite{201903tvt}. On the receiver side, BS and UE nodes can perform interference cancellation using pre-trained CSI~\cite{201810tvt}. However, this method introduces a high cost of implementation and requires CSI for all possible CLI channels. As demonstrated by the dynamic TDD scenario, CLI handling is essential for all future mobile network models (including dynamic TDD, flexible duplex, and FD) that support asymmetric and time-varying demands for UL and DL transmissions.

\subsection{Management Strategies Against CLI in Full-duplex Networks}
		
	FD communications achieve higher flexibility through more aggressive resource utilization than the dynamic TDD system. Enhanced CLI mitigation techniques and the SIC are necessary to make this possible. In~\cite{202102wcom2}, the simulation results showed that the FD-based network potentially has a higher DL and UL capacity than the dynamic TDD but is also limited by severer CLI. Moreover, CLI often served as a bottleneck for the system, making the adoption of strategies to avoid CLI highly desirable in FD network scenarios~\cite{201505commag,2022icc}. 

{In order to mitigate CLI, network nodes require channel estimation and additional algorithms, which brings hardware and software overheads~\cite{201505commag1}. }
The impact of the CLI becomes more severe than in dynamic TDD systems due to several factors. First, since the UE nodes are mostly assumed to operate in HD mode, all time-frequency resources can have two different UE nodes for UL and DL links. 
Second, intra-cell UE-to-UE interference exists, leading to the study of intra-cell from UL node to DL node interference for single-cell FD operation~\cite{201406cl}. To mitigate CLI, two primary steps should be taken: first, a CLI measurement scheme, which is essential for link scheduling and post-processing; and then, the implementation of enhanced algorithms and appropriate network design to manage and minimize interference.


\subsubsection{CLI Measurement}
	
	
Channel measurement for CLI can be utilized for various mitigation techniques from the network level. CLI can be pre-avoided through system design or coordinated operation of BSs. For this purpose, the power level of CLIs at the cell BS nodes must be distributed to other cells. BS-to-BS CLI measurement is straightforward and should be avoided at the network level. However, utilizing UE-to-UE CLI channel information requires additional protocols for the network. CLI measurement and reporting for the coordination of BSs have been actively discussed in the literature. In the FD networks, standardized measurement methods for the dynamic TDD system can be utilized~\cite{202102wcom}.

	Two different channel estimation schemes can be considered for the centralized coordination of BS and UE nodes to mitigate the UE-to-UE CLI~\cite{ts38215}. In the CLI-sounding reference signal (CLI-SRS) method, the gNB configures a victim UE node and one or more possible interferers through RRC signaling. At the configured SRS resources, a victim UE measures the reference signal received power (RSRP) of the transmitted SRS of the interfering UE nodes.
	In the CLI-received signal strength indicator (CLI-RSSI) method, a victim UE measures the RSSI of the configured resources to estimate the power level of the CLI for possible DL reception. The estimated CLI channel is then transmitted back to the gNB through RRC signaling. The configured UE nodes can report the measurements periodically or triggered by a specific condition~\cite{ts38331}. By utilizing the measurement of the CLI channel information, various techniques can be jointly conducted by multiple gNBs to mitigate the impact of CLI in FD networks.

	\subsubsection{CLI mitigation techniques for cellular FD network}

	In a cell-based FD network, the type of interference of most interest is the intra-cell UE-to-UE CLI. Because the DL and UL users can occupy the same sub-band within the cell, UE-to-UE CLI can be significant depending on their specific location and CLI channel. In order to avoid system overhead at the BS for CLI channel estimation and site-specific user scheduling, simple post-processing at the users can be assumed. In~\cite{201603ietc,201710tvt}, a single-cell scenario with a BS operating in FD mode and users operating in HD mode was assumed. The BS could first prevent UE-to-UE interference through spatial precoding and resource allocation strategies. Then, residual UE-to-UE CLI was removed from the received signal by decoding the CLI signal at each user. Unlike previous studies, in~\cite{201904twc}, the assumption of perfect channel information at the BS node was tackled. The CSI of the UE-to-UE CLI was estimated through the UL pilot signal and utilized for the CLI cancellation. The performance of the algorithm was optimized by solving the power allocation problem for the pilot and data symbols. 
	
	Interference cancellation is the most intuitive and effective method if the cancellation can be done perfectly. However, it requires a high computation complexity, and the performance of the cancellation depends on the channel estimation accuracy. Without interference cancellation at the receiver, various strategies had been proposed to minimize the impact of CLI. In~\cite{201406cl}, the authors proposed an algorithm to sectorize the cell region with different sub-channels allocated. The UE-to-UE CLI could be minimized by maximizing the distance between the UL and DL transmission sectors using the same sub-channel. Furthermore, a game-theoretic user allocation strategy was proposed in~\cite{202112access}. The algorithm was based on user competition, like the concept of contention in FD WLAN, for sub-channels where the operator only allowed an allocation when the sum utility, which is a function of SINR, increased.

	A single-cell scenario is mostly assumed where the UE-to-UE CLI is critical. However, some studies proposed mitigation techniques for the BS-to-BS CLI in multi-cell scenarios. Researchers conducted experiments and system-level simulations for mmWave FD BSs with three sectors~\cite{2022icc}. Two $\mathtt{Tx}$ antenna arrays and two $\mathtt{Rx}$ antenna arrays were installed in each sector, with sectors facing different directions. Throughout the simulation of the multi-cell scenario, the authors prevented CLI by employing narrow mmWave beamforming and showed that the SINR loss was minimal compared to the HD counterpart when SI between the sectors was well canceled. In~\cite{202212access}, the authors proposed SIC and CLI management processes for a sub-band FD system. The BS-to-BS CLI was mitigated through interference identification and power allocation to manage the interference level within the cell.

	\subsubsection{Evolution of FD network models for efficient CLI handling}

	Conventional approaches for dynamic TDD and FD are mostly implemented in cellular networks, where FD-enabled BSs independently service cell users. The BS nodes mitigate inter-user interference and intra-cell UE-to-UE CLI through proper techniques. To mitigate inter-cell CLIs, coordinated channel estimation and resource allocation of nearby BSs are often assumed~\cite{202212access,2015icc,2021vtc}. However, interference mitigation techniques for multi-cell scenarios are impractical due to the complexity and variety of interference types, even though the potential performance enhancement of FD-based communications was proved. Therefore, advancements were made in network models to handle CLI with reasonable network complexity.
	
\begin{figure*}

	\begin{center}
		{\includegraphics[width=2\columnwidth,keepaspectratio]
			{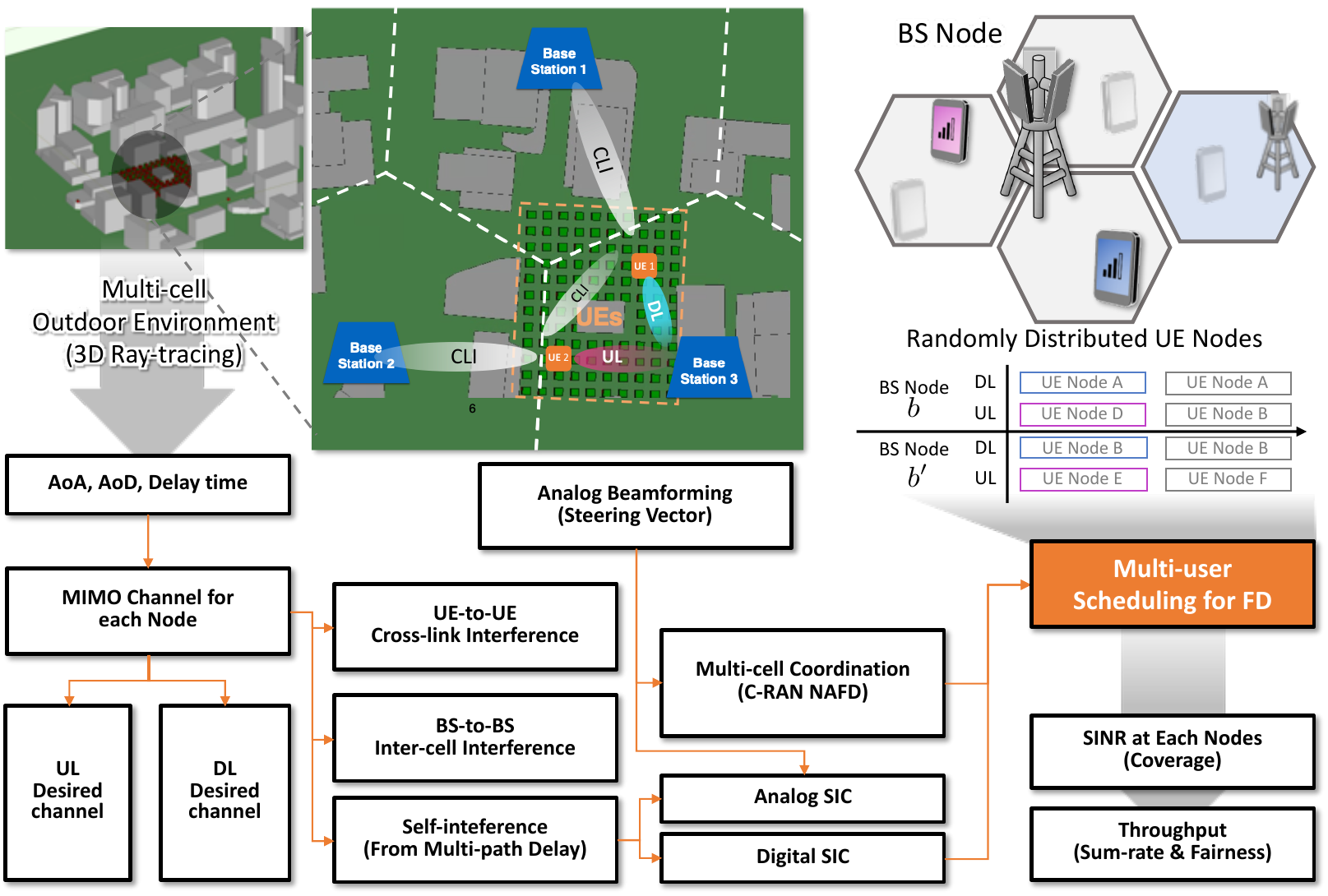}%
			\caption{Evaluation method based on 3D ray-tracing channel realization.}
			\label{fig.eval}
		}
	\end{center}
	
\end{figure*}

	The FD network enabled by CoMP was proposed in~\cite{201604wcl}, referred to as CoMPflex. The $\mathtt{Tx}$ and $\mathtt{Rx}$ antennas of BS nodes were spatially separated to reduce the overhead of SIC, but they acted like a single BS through the fiber connection. The authors analyzed the performance of IBFD service on UL and DL users for varying distances between the two HD BS nodes. This study demonstrated that the distribution of APs brought advantages regarding interference mitigation.
	
	Inspired by the C-RAN architecture, numerous network models with distributed antennas and centralized processors were studied~\cite{201408wcl,201501cst}. A network model of the distributed antenna system (DAS) for the bidirectional dynamic network (BDN) was analyzed in~\cite{201703ietc}. The network consisted of a single baseband unit and multiple remote radio heads (RRHs), which were antenna arrays distributed in the network area. The authors solved the data traffic asymmetry problem by adjusting the number of UL and DL RRHs. Furthermore, the analytical and numerical results showed that inter-RRH CLI vanished as the number of antennas of the massive MIMO system increased. The BDN implemented by DAS had a primary advantage in jointly servicing massive users and preventing SI. Moreover, the centralized operator enabled handling UE-to-UE CLI by the centralized operation of the baseband unit~\cite{201703access,201807access,201809twc}. The BDN operated as an in-band FD network by servicing UL and DL users in the same frequency band; however, the flexibility of the FD network was still limited by the HD operation of each RRH.

	The FD network model with full flexibility was proposed in~\cite{202003tcom}. The NAFD was based on cell-free massive MIMO with distributed remote antenna units (RAUs) and a central processing unit (CPU), similar to the RRHs and the baseband unit of the DAS-based BDN, respectively. The difference between NAFD and BDN was that all transmission modes, including UL/DL only, flexible duplex, and FD, were allowed for each RAU in the NAFD. Thus, the CPU could optimize the total utility of the network by allocating different UL/DL configurations to the RAUs. Various techniques for handling CLI in NAFD networks were proposed in~\cite{202003tcom,202008jsac,2023eajsac,202104twc}.

	First, pre-processing at the CPU was mainly considered to handle UE-to-UE CLI since the CLI cancellation method imposed a high implementation cost to UEs. The CPU could adopt enhanced transmission strategies to minimize the impact of UL-UE node transmission on DL-UE node reception. In~\cite{202008jsac}, UL-UE nodes transmitted with zero-forcing (ZF) beamforming to minimize the interfering signal directed toward other DL-UE nodes. Also, appropriate user scheduling algorithms that consider the interference channel between users could significantly reduce the impact of CLI~\cite{202003tcom}. By exploiting the high flexibility of the NAFD network, the authors in~\cite{2023eajsac} proposed to mitigate UE-to-UE CLI by joint RAU mode selection and power control.

	Due to the higher computation capacity and quasi-static nature of the BS-to-BS CLI channel, it can be a practical assumption to cancel out CLI at the UL-RAUs. In~\cite{202003tcom}, BS-to-BS CLI was canceled in the digital domain using pre-trained CSI, while SIC was implemented in the analog domain. In~\cite{202008jsac}, the advanced precoder developed by the principal component analysis (PCA) of the CLI channel was applied at the DL-RAUs to avoid generating a transmitting beam directed towards the UL-RAUs. However, an imperfect CSI of the BS-to-BS CLI limited the effect of mitigation techniques and deteriorated the system performance~\cite{201807access,201904twc,202003tcom}. The acquisition of the BS-to-BS CSI is fundamental but also a challenging process for implementing centralized CLI mitigation algorithms. In~\cite{202104twc}, a novel CSI acquisition method was developed for BS-to-BS CLI cancellation and DL transmissions. The authors in~\cite{202008jsac} showed that the proposed heap-based pilot assignment algorithm achieved higher channel estimation accuracy, further enhancing the average spectral efficiency and energy efficiency of the system.
	
\begin{figure*}

	\begin{center}
		\subfigure[Downlink, at DL-UE node.]{\includegraphics[width=1\columnwidth,keepaspectratio]
			{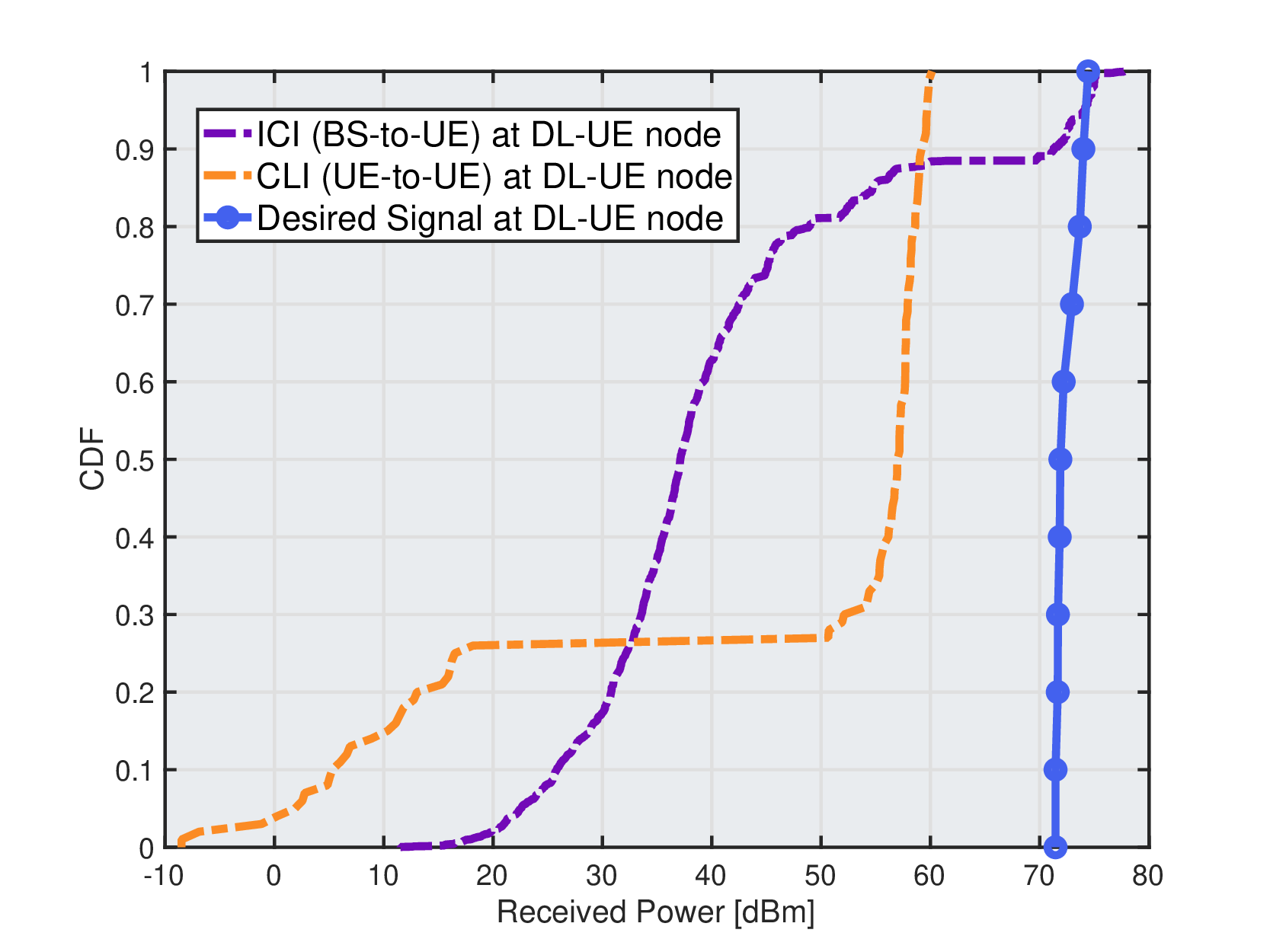}%
			\label{fig.INIdl}
			}
		\subfigure[Uplink, at BS node.]{\includegraphics[width=1\columnwidth,keepaspectratio]
			{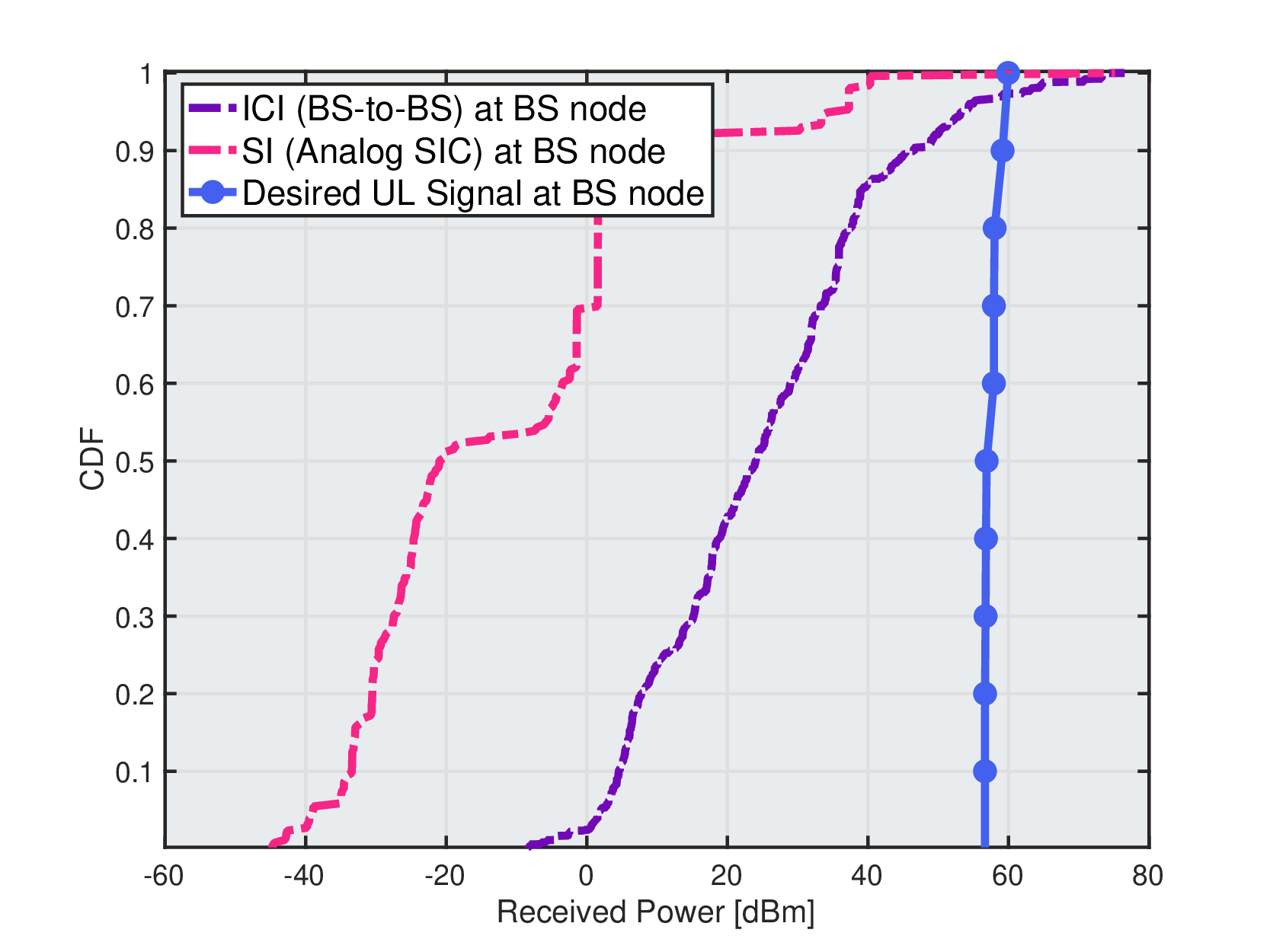}%
			\label{fig.INIul}
			}
			
	\caption{Inter-node interference (INI) measurement based on 3D ray-tracing. The propagation environment is multi-path scenario, and each INI occurs through multi-path channel. The analog $\mathtt{T/Rx}$ beam is selected to maximize desired channel gain.}
	\label{fig.INI}
	\end{center}
	
\end{figure*}

\begin{center}
	\begin{table}[t] 
		\caption{Simulation Parameters}
		\begin{tabular}{|>{\centering } m{1.7cm} |>{\centering} m{3.9cm} |>{\centering} m{1.9cm} | }
			\hline
			\textbf{Paramet{\tiny }er} & \textbf{Description} & \textbf{Value}
			\tabularnewline
			\hline
			\multicolumn{3}{|c|}{\textbf{Sub-6 GHz channel parameters for 3D ray-tracing}}
			\tabularnewline\hline
			\centering		$f_\mathrm{c}$  & Carrier frequency & 3.5 (GHz)  \tabularnewline \hline
			\centering		$\text{BW}$  & Bandwidth & 100 (MHz)  \tabularnewline \hline
			\centering		$N_\text{Ray}$ & Number of rays to consider multi-path channel& 25 \tabularnewline \hline
			\multicolumn{3}{|c|}{\textbf{BS node (FD-compatible) specifications}} 
			\tabularnewline\hline
			\centering		$N_\text{BS}$  & Number of BS nodes (cells) & 3  \tabularnewline \hline
			\centering		$M^\text{BS}_\mathtt{Tx}, M^\text{BS}_\mathtt{Rx}$  & Antenna array  & $4 \times 4~(16)$ UPA \tabularnewline \hline
			\centering		$P=|\mathbb{P}|$  & Number of subcarriers& 2048 \tabularnewline \hline
			\centering		$\delta_\text{ISO}$ & Isolation between $\mathtt{Tx}$ and $\mathtt{Rx}$ & 0.5(m) \tabularnewline \hline
			\centering          $h_\text{BS}$ & BS node height & $20$~(m)\tabularnewline \hline
			\centering		$\sigma^2$  & Noise figure & 10 (dB)  \tabularnewline \hline
			\centering		$p^\text{DL}_b$  & Transmit power of BS nodes & $[0, 50]$ (dBm)  \tabularnewline \hline
			\centering		$G^\text{SI,digital}_b$  & Digital SIC level & $[-20, -50]$~(dB)  \tabularnewline \hline
			\multicolumn{3}{|c|}{\textbf{UE node (HD) specifications}} 
			\tabularnewline\hline
			\centering		$N_\text{UE}$  & Number of UE nodes & 20  \tabularnewline \hline
			\centering		$M^\text{UE}_\mathtt{Tx}, M^\text{UE}_\mathtt{Rx}$  & Antenna array  & $4 \times 4$ UPA  \tabularnewline \hline
			\centering		$h_\text{UE}$  & UE node height & $1.5$~(m)  \tabularnewline \hline
			\centering		$p^\text{UL}_i$  & Transmit power of UE nodes& $15$~(dBm)  \tabularnewline \hline
		\end{tabular}
		\label{table.SimPar}
	\end{table}
\end{center}
\section{3D Ray-tracing-based \\Full-duplex Network Evaluation}\label{sec.EVAL}
\subsection{System-level Simulator Setup}

Our objective is to examine the impact of scheduling algorithms and INIs on FD network design. We carried out a SLS similar to a real-world multi-cell communications environment, as shown in Fig.~\ref{fig.eval}. To do this, we used 3D ray-tracing to model the channel in an outdoor multi-cell scenario and conducted a comparison of scheduling techniques.

Although numerous scheduling algorithm researches exist, only~\cite{US-NAFD} has considered MIMO beamforming compatible base BS nodes. Given that modern communication systems and standard BS technologies rely on multiple antennas, it is essential to evaluate the impact of existing scheduling in the context of MIMO. In addition to the well-explored impact of SIC, this FD network SLS aims to address the factors that have been overlooked in previous FD network studies

To assess the impact of various INIs in multi-cell networks, multiple duplex and scheduling options were evaluated through system-level simulation, taking each INI into account. SI, where suppression is most expected; ICI, which can be anticipated to be measured through coordination; and CLI, which requires complex measurement at the UE node level, were all considered in the network optimization.

To obtain the actual channels for multiple users, we measured MIMO channels between nodes through 3D ray-tracing. We generated desired UL and DL channels at various points, a BS-to-BS channel to reflect the multi-cell environment and a channel between users. UL- and DL-UE nodes become user pairings for each RB. We performed ray-tracing in an outdoor environment, verifying the distribution of each type of INI, as shown in Fig.~\ref{fig.INI}.

In exploring the effect of the transmit power disparity, $\delta$, discussed in Section~\ref{sec.US}, we set the transmit power of the UE node to a constant $15$~dBm, allowing the transmit power of the BS node to vary. We ensured fairness for each BS node by assigning a minimum of a single time slot to all UE nodes. We considered both perfect scheduling and simple round-robin scheduling in our analysis. Table~\ref{table.SimPar} provides a detailed account of the parameters used.

\subsection{Simulation Results}

\begin{figure}
	\begin{center}
		{\includegraphics[width=1\columnwidth,keepaspectratio]
			{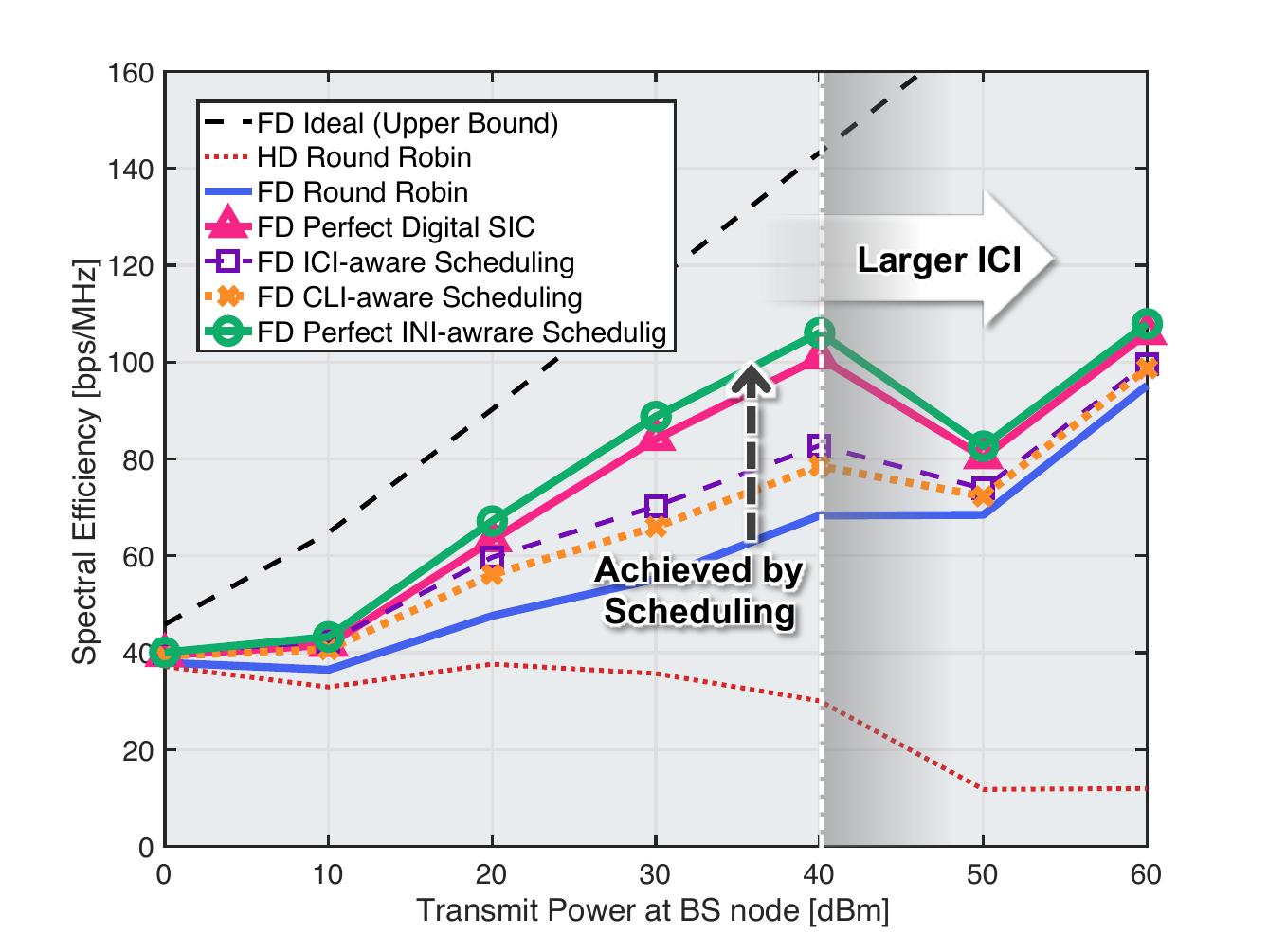}%
			\caption{Evaluation result based on 3D ray-tracing channel realization.}
			\label{fig.res}
		}
	\end{center}
	
\end{figure}

Fig.~\ref{fig.res} presents the spectral efficiency across the network based on the degree of provided or estimated INI. We achieved optimal performance with FD perfect scheduling, which accounts for all INIs in the scheduling process. Among the three components of INI-SI, ICI, and CLI-SIC technology, that eliminates SI from the BS node exerts the most substantial influence. It appears that ICI and CLI have a similar level of consideration in the scheduling process.

\subsubsection{Impact of SI and digital SIC}
In FD networks, SIC, a significant component of link-level FD, greatly influences performance. As Fig.~\ref{fig.INIdl} demonstrates, the size of SI varies due to different pairing combinations depending on the location of the UE node. The digital SIC employed by each node is designed to reduce SI to the noise level. As a result, implementing perfect SIC helps streamline the optimization process by ensuring consistent residual SI from a network optimization perspective. With digital SIC in place, the scheduling algorithm can prioritize optimizing user pairing to minimize CLI and ICI.

\subsubsection{Impact of ICI by BS coordination}
ICI significantly impacts the entire FD network. As shown in Fig.~\ref{fig.res}, when the transmit power at the BS node exceeds $40$~dBm, all FD performance fluctuates due to the increased ICI from the adjacent BS node. Performance fluctuations continue even with perfect SIC and optimized scheduling, highlighting the need for ICI coordination research for interference suppression. 

\subsubsection{Impact of CLI by user scheduling}
Although CLI is the most challenging INI to measure, we can address it through scheduling due to its site-specific nature in environments with nodes. As evident in Fig.~\ref{fig.INIdl}, the size of CLI exceeds that of ICI, which has the most significant impact on the FD network. However, unlike ICI, the network can reduce the impact of CLI through proper user pairing, underlining the need for research into user scheduling, especially regarding beamforming.

Future research may explore scheduling by integrating OFDMA and MIMO concepts. The increased flexibility will minimize throughput, fairness, and cell coverage tradeoffs.

\section{Conclusion}\label{sec.CNCL}
This article presented an overview of the latest research on full-duplex (FD) network design. Applying FD to various networks, such as WLAN and cellular networks, poses a significant challenge due to various types of inter-node interference. Therefore, we provided a comprehensive analysis of the strategies to tackle these inter-node interference problems from a network perspective across the MAC/RRC/PHY layers.


We covered MAC protocols for establishing FD links, multi-cell user scheduling algorithm, and how to measure and handle cross-link interference from the 3GPP document from a PHY perspective. In particular, network scheduling has been studied very actively and can be combined with the cell-free network and cloud RAN, which are the recent concept of future networks. We conducted system-level simulations based on 3D ray-tracing to verify the primary considerations of FD network design and the impact of the summarized technologies in a practical multi-cell environment.

Designing an FD network necessitates the consideration of inter-node interference and the fairness of servicing of nodes. Future works related to FD network design need to manage interference between nodes at the network level from a protocol perspective, following the recent transceiver structures, such as MIMO. Furthermore, the algorithms need to be simplified to enable implementation with low complexity and low latency, advancing beyond the current optimization stage in academic research.   


\bibliographystyle{IEEEtran}
\bibliography{Ref_PROC}	

%


\vfill


\begin{IEEEbiography}[{\includegraphics[width=1in,height=1.25in,clip,keepaspectratio]{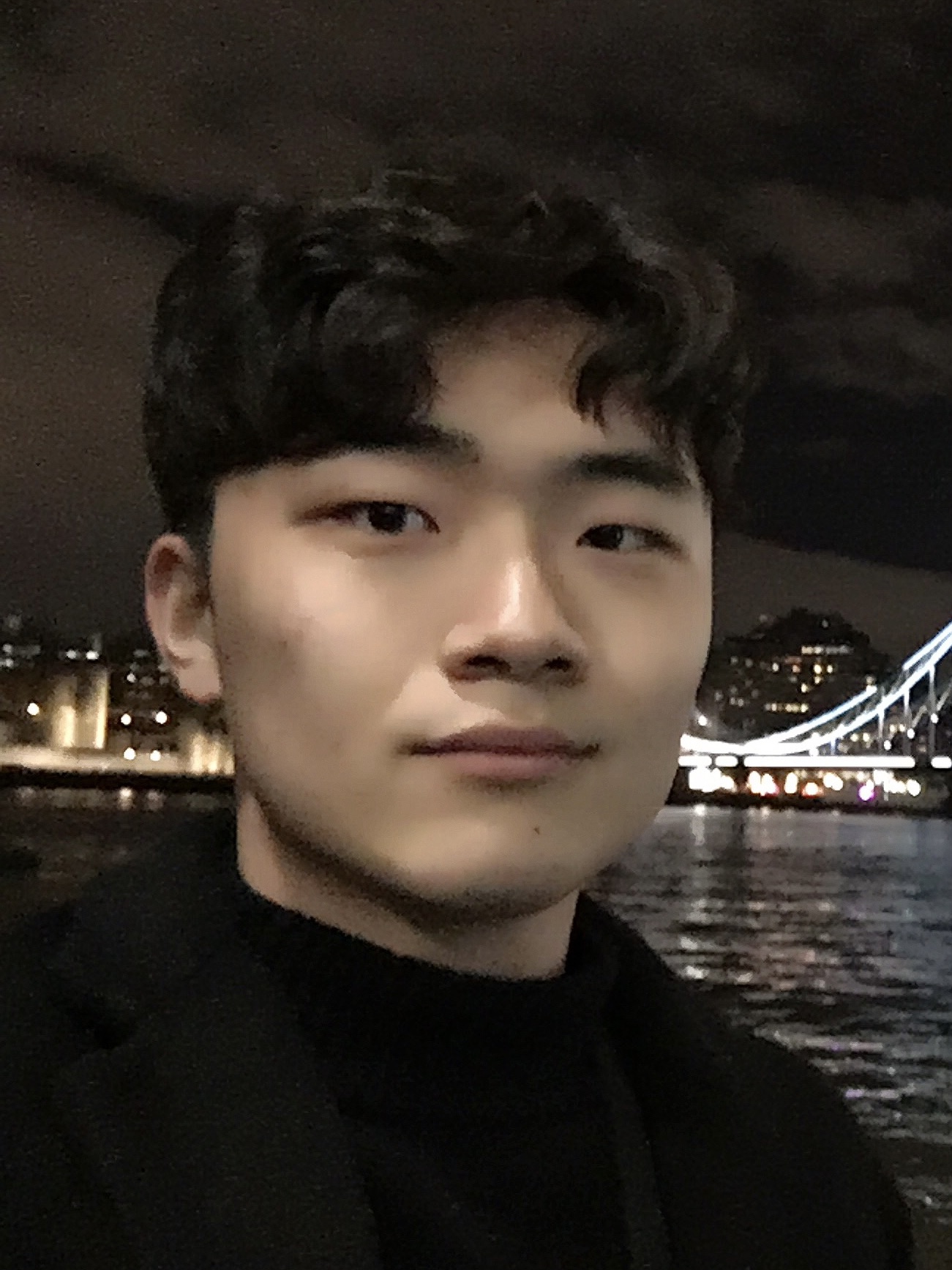}}]{Yonghwi Kim} (Student Member, IEEE) received the B.S. degree from the School of Integrated Technology, Yonsei University, South Korea, in 2019, where he is currently pursuing the Ph.D. degree. His research interest includes resource allocation problems for B5G/6G communications such as full-duplex (FD) radios, reconfigurable intelligent surface (RIS), and integrated sensing and communications (ISAC).
\end{IEEEbiography}

\begin{IEEEbiography}[{\includegraphics[width=1in,height=1.25in,clip,keepaspectratio]{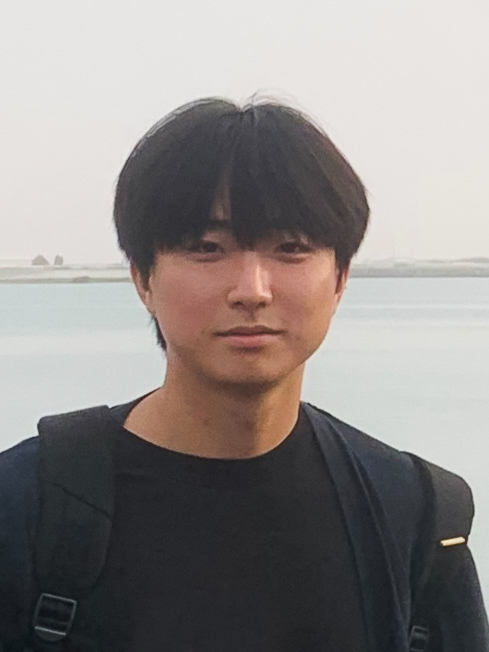}}]{Hyung-Joo Moon} (Student Member, IEEE) received the B.S. degree from the School of Integrated Technology, Yonsei University, South Korea, in 2019, where he is currently pursuing the Ph.D. degree. His research interests include performance analysis and system optimization for emerging technologies in 6G non-terrestrial networks (NTNs).
\end{IEEEbiography}

\begin{IEEEbiography}[{\includegraphics[width=1in,height=1.25in,clip,keepaspectratio]{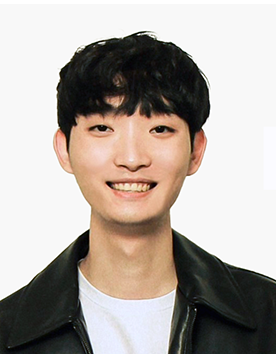}}]{Hanju Yoo} (Student Member, IEEE) received the B.S. degree (Summa Cum Laude, ranked 1st) from the School of Integrated Technology, Yonsei University, South Korea, in 2021, where he is currently pursuing his Ph.D. degree. His research interests include deep neural networks for computer vision, learned image compression, semantic communications/deep joint source-channel coding, and Vision Transformer architecture.
\end{IEEEbiography}

\begin{IEEEbiography}
[{\includegraphics[width=1in,height=1.25in,clip,keepaspectratio]{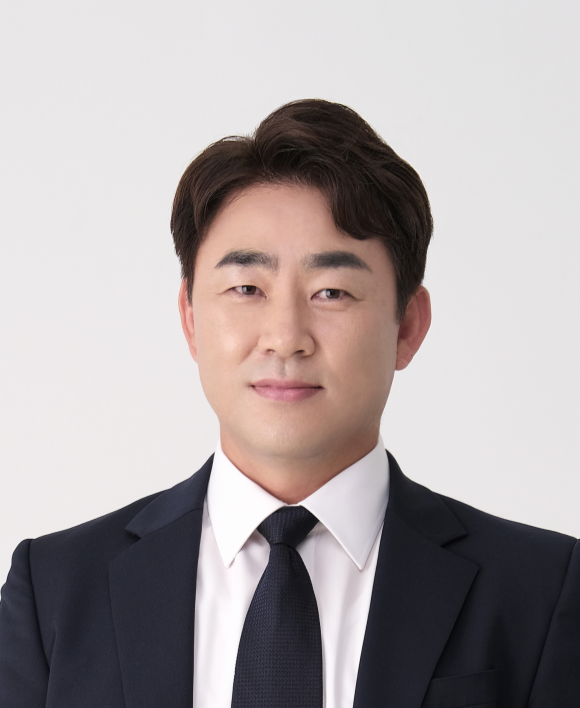}}]{Byoungnam (Klaus) Kim} (Member, IEEE) earned his Ph.D. in Electrical Engineering from the Korea Advanced Institute of Science and Technology (KAIST) in South Korea. He currently serves as the CEO of SensorView Ltd, a company that provides a competitive edge in the next generation of 5G technologies and IoT mobile communications using unique materials and product design technologies. Prior to founding SensorView in 2015, he held leadership roles as Head of Business Unit and Head of R\&D at Ace Technologies. His research interests focus on advanced RF and communications technologies for 5G/B5G/6G.

Dr. Kim has received multiple esteemed accolades throughout his career, including the CES Innovation Awards in 2023 and recognition from the Ministry of SMEs and Startups in 2019, the Ministry of Knowledge Economy in 2011, and the Governor Award of Gyeonggi-do in 2020. SensorView, the company he founded, has been honored as a Global Unicorn by the Ministry of Science and ICT in 2023 and as a Pre-Unicorn by the Ministry of SMEs and Startups in 2022. SensorView was also recognized as one of the 15 outstanding industrial technology achievements by the National Academy of Engineering of Korea (NAEK) in 2021.
\end{IEEEbiography}

\begin{IEEEbiography}[{\includegraphics[width=1in,height=1.25in,clip,keepaspectratio]{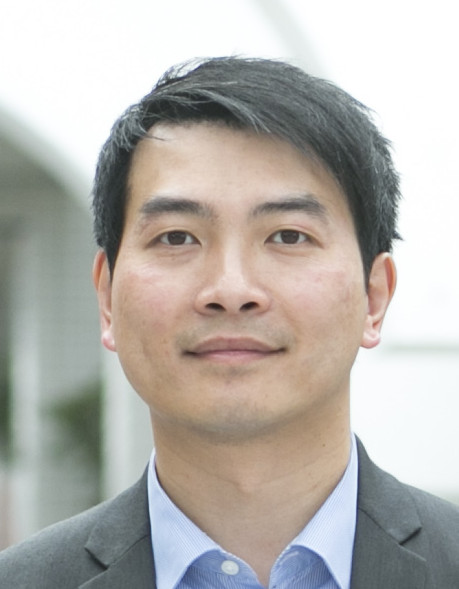}}]{Kai-Kit Wong} (Fellow, IEEE) received the BEng, the MPhil, and the PhD degrees, all in Electrical and Electronic Engineering, from the Hong Kong University of Science and Technology, Hong Kong, in 1996, 1998, and 2001, respectively. After graduation, he took up academic and research positions at the University of Hong Kong, Lucent Technologies, Bell-Labs, Holmdel, the Smart Antennas Research Group of Stanford University, and the University of Hull, UK. He is Chair in Wireless Communications at the Department of Electronic and Electrical Engineering, University College London, UK. His current research centers around 5G and beyond mobile communications. He is a co-recipient of the 2013 IEEE Signal Processing Letters Best Paper Award and the 2000 IEEE VTS Japan Chapter Award at the IEEE Vehicular Technology Conference in Japan in 2000, and a few other international best paper awards. He is Fellow of IEEE and IET and is also on the editorial board of several international journals. He is the Editor-in-Chief for IEEE Wireless Communications Letters since 2020.

\end{IEEEbiography}

\begin{IEEEbiography}[{\includegraphics[width=1in,height=1.25in,clip,keepaspectratio]{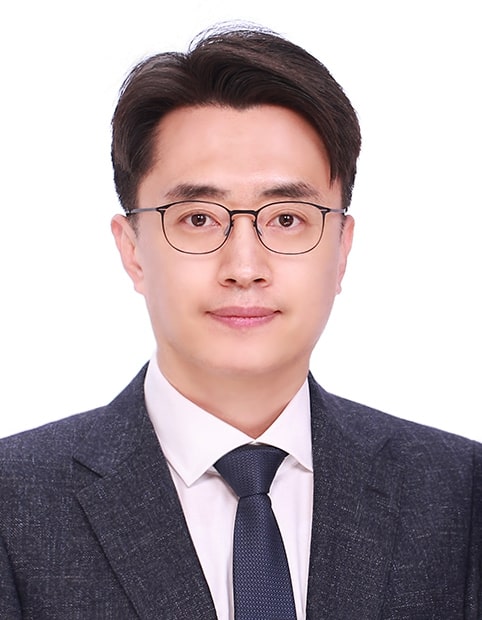}}]{Chan-Byoung Chae} (Fellow, IEEE) received the Ph.D. degree in electrical and computer engineering from The University of Texas at Austin (UT), USA in 2008.

Prior to joining UT, he was a Research Engineer at the Telecommunications Research and Development Center, Samsung Electronics, Suwon, South Korea, from 2001 to 2005. He is currently an Underwood Distinguished Professor with the School of Integrated Technology, Yonsei University, South Korea. Before joining Yonsei University, he was with Bell Labs, Alcatel-Lucent, Murray Hill, NJ, USA, from 2009 to 2011, as a Member of Technical Staff, and Harvard University, Cambridge, MA, USA, from 2008 to 2009, as a Post-Doctoral Research Fellow.

Dr. Chae was a recipient/co-recipient of the KICS Haedong Scholar Award in 2023, the CES Innovation Award in 2023, the IEEE ICC Best Demo Award in 2022, the IEEE WCNC Best Demo Award in 2020, the Best Young Engineer Award from the National Academy of Engineering of Korea (NAEK) in 2019, the IEEE DySPAN Best Demo Award in 2018, the IEEE/KICS Journal of Communications and Networks Best Paper Award in 2018, the IEEE INFOCOM Best Demo Award in 2015, the IEIE/IEEE Joint Award for Young IT Engineer of the Year in 2014, the KICS Haedong Young Scholar Award in 2013, the \textit{IEEE Signal Processing Magazine} Best Paper Award in 2013, the IEEE ComSoc AP Outstanding Young Researcher Award in 2012, and the IEEE VTS Dan. E. Noble Fellowship Award in 2008. 

Dr. Chae has held several editorial positions, including Editor-in-Chief of the \textsc{IEEE Transactions on Molecular, Biological, and Multi-Scale Communications}, Senior Editor of the \textsc{IEEE Wireless Communications Letters}, and Editor for the \textit{IEEE Communications Magazine}, \textsc{IEEE Transactions on Wireless Communications}, and \textsc{IEEE Wireless Communications Letters}. He is an IEEE ComSoc Distinguished Lecturer from 2020 to 2023. He is an IEEE Fellow and NAEK Fellow.
\end{IEEEbiography}

\end{document}